\documentclass[sigconf, nonacm]{acmart}

\usepackage{verbatim}
\usepackage{amsmath}
\usepackage{tikz}
\usepackage{subfigure}
\usepackage[ruled,vlined]{algorithm2e}
\usepackage{balance}
\usepackage[T1]{fontenc}

\usepackage{titlesec}
\titlespacing*{\section}{0pt}{6pt plus 2pt minus 1pt}{4pt plus 1pt minus 1pt}
\titlespacing*{\subsection}{0pt}{4pt plus 2pt minus 1pt}{2pt plus 1pt minus 1pt}
\titlespacing*{\subsubsection}{0pt}{3pt plus 1pt minus 1pt}{2pt plus 1pt minus 1pt}
\titlespacing*{\myparagraph}{0pt}{3pt plus 1pt minus 1pt}{2pt plus 1pt minus 1pt}

\usepackage{caption}
\newcommand{\myparagraph}[1]{
\noindent{\textbf{#1}}}
\newcommand{\certa}[0]{\textsc{certa}}
\newcommand{\uncerta}[0]{\textsc{Ellmer}}

\newcommand{\lemon}[0]{\textsc{Lemon}}
\newcommand{\minun}[0]{\textsc{Minun}}
\usepackage{balance}
\usepackage{paralist, tabularx}
\usepackage{enumitem}

\begin{document}

\title{Can we trust LLM Self-Explanations for Entity Resolution? 
}

\author{Tommaso Teofili}
\affiliation{%
  \institution{Roma Tre University, Elastic}
}
\email{tommaso.teofili@uniroma3.it}

\author{Donatella Firmani}
\affiliation{%
  \institution{Sapienza University}
}
\email{donatella.firmani@uniroma1.it}

\author{Nick Koudas}
\affiliation{%
  \institution{University of Toronto}
}
\email{koudas@cs.toronto.edu}

\author{Paolo Merialdo}
\affiliation{%
  \institution{Roma Tre University}
}
\email{paolo.merialdo@uniroma3.it}

\author{Divesh Srivastava}
\affiliation{%
  \institution{AT\&T Chief Data Office}
}
\email{divesh@research.att.com}

\begin{abstract}
Large Language Models (LLMs) have recently shown strong performance on Entity Resolution (ER). Additionally, akin to their prowess in providing accurate predictions, these models often generate self-explanations alongside their predictions through prompting. While such self-explanations are appealing due to their negligible computational cost, their actual reliability remains largely unexplored.
In this paper, we present the first large-scale systematic evaluation of LLM self-explanations for ER, focusing on feature attribution and counterfactual explanations at both the attribute and token levels. Across three LLMs, ten datasets, and multiple prompting strategies, we show that self-explanations are often unstable, weakly faithful, and poorly aligned with counterfactual evidence, revealing a substantial gap between plausibility and causal relevance.
We further demonstrate that established post-hoc explanation methods provide significantly higher trustworthiness, but at a prohibitive computational cost when applied to LLMs. To bridge this gap, we introduce \uncerta{}, a hybrid explanation framework that leverages self-explanations as priors to guide post-hoc exploration. \uncerta{} achieves explanation quality comparable to post-hoc methods while reducing cost by up to an order of magnitude.
%
\end{abstract}

\maketitle

\section{Introduction}
\label{sec:intro}

Entity Resolution (ER) is the fundamental problem of matching different representations of the same entity – e.g., a book or a customer record – across heterogeneous data sources, such as different
review websites or corporate databases. As the usage of large, diverse datasets grows, ER has become an essential step in data preparation, enabling the integration and creation of new business-ready datasets.
Recent advances in deep learning have been leveraged to tackle the ER problem, leading to significant improvements in accuracy and performance~\cite{ditto,peeters_2021,zeakis2023pre,zecchini2023brewer,guo2023camper,li2024booster,fan2024cost,aberbach2024multipartite,paganelli2024multi,zeakis2025depth}. However, these techniques rely on large volumes of training data, specifically labeled pairs of records, where each pair is annotated as either a match or a non-match.  

Given the demonstrated power of Large Language Models (LLMs) and their considerable success across various domains -- particularly in zero-shot and few-shot learning -- recent studies have investigated their potential to tackle the ER problem, producing impressive results~\cite{DBLP:conf/adbis/PeetersB23,sisaengsuwanchai2023does,xia2024aprompt4em,li2024booster,nananukul2024cost,wang2025match}.

Traditional ER systems based on deep learning work as closed models whose inner working mechanisms are opaque, and they can sometime make the right decisions for the wrong reasons~\cite{mojito,thirumuruganathan2019explaining}. 
To overcome this limitation, several post-hoc methods have been proposed to generate explanations for the predictions of such ER systems~\cite{mojito,baraldi2021landmark,baraldi2023intrinsically,minun,teofili2022effective,teofili2022certem,lemon,benassi2024explaining}. 

LLMs are also opaque. However, along with their ability to produce precise predictions, they exhibit a unique trait: the ability of generating \emph{self-explanations}~\cite{madsen-etal-2024-self} together with their ER outcomes.

Our work embarks on a thorough investigation to assess the quality and scalability of self-explanations and post-hoc explanations for ER predictions generated by LLMs. In particular, we analyze \emph{feature attribution} and \emph{counterfactual} explanations~\cite{10.1145/3461702.3462597,10.1145/3546577}, which are widely used in ER systems~\cite{10.1145/3328519.3329130,jesus2021can} and exhibit high usability~\cite{wang2021explanations,rong2023towards}. The former assigns an \emph{attribution score} to each feature in the input pair, indicating its contribution to the prediction; the latter provides examples of value changes that can alter the prediction.

In particular, we report the results of experimental evaluations which aim to examine three key research questions (RQ):

\myparagraph{RQ-1: Are self-explanations trustworthy?} 

Prior work has shown that LLM outputs may exhibit sensitivity to prompt variations across tasks~\cite{lu-etal-2022-fantastically,zhao2021calibrate,zhuo2024prosa}, raising concerns about the stability of their generated explanations.

We therefore assess the \emph{robustness of the self-explanations} for the ER task under different prompting strategies.
A further concern is the occurrence of ``hallucinations'', which in our context refer to the generation of self-explanations that appear plausible but are not grounded in the actual behavior of the LLM. 
Regrettably, such undesired behaviors could undermine the \emph{reliability of self-explanations}, thus hampering the effectiveness of ER predictions with self-explanations. 
To assess this issue, we conduct an experimental evaluation using established metrics from the explainable AI literature: faithfulness~\cite{saliency_metrics} for feature attribution, and validity~\cite{cf_metrics} for counterfactual self-explanations. While faithfulness and validity do not capture all dimensions of explanation quality, they are widely adopted proxies for assessing whether explanations are grounded in the actual decision logic of a model~\cite{jacovi2020towards}. Also, we assess the consistency between attribute level and token level explanations, and the agreement between feature attribution and counterfactual explanations. 

\myparagraph{RQ-2: Are post-hoc explanations more trustworthy?}
Given the availability of post-hoc explanations for black-box ER systems, we wonder whether they are more reliable than self-explanations. To this end, our investigation proceeds comparing the self-explanations of LLMs to the post-hoc explanations provided by  systems specifically developed for ER. 
Our comparison encompasses both feature attribution and counterfactual explanations, and involves the SOTA systems {\certa}~\cite{teofili2022effective,teofili2022certem}, \lemon{}~\cite{lemon}, and \minun{}~\cite{minun}. Notably, all these methods are model-agnostic and thus can be applied to LLMs as well. 

\myparagraph{RQ-3: Is it feasible to compute post-hoc explanations for LLMs?} Our experiments unveil an intriguing trade-off: 
post-hoc explanations exhibit significantly higher reliability compared to self-explanations. However, this reliability comes at a computational cost, as post-hoc methods require additional processing steps.
To address the trade-off between reliability and efficiency, we propose {\uncerta}, a hybrid explanation method that seamlessly integrates the inherent efficiency of self-explanations with the trustworthiness of post-hoc explanations to enhance reliability. This approach not only mitigates the risk of hallucinations but also ensures scalability, making it a practical solution for real-world ER
applications.

To address the above research questions, we conduct an experimental evaluation using three LLMs and ten datasets, 
allowing us to assess the robustness and generality of the proposed analysis across heterogeneous settings.

In summary, this paper makes the following contributions:
\begin{itemize}[leftmargin=*]
\item We systematically evaluate the trustworthiness of LLM-generated self-explanations for entity resolution, analyzing their sensitivity to prompting strategies and their susceptibility to hallucinations.
\item We compare self-explanations and post-hoc explanations in terms of both effectiveness and computational efficiency in the ER setting.
\item We experimentally assess a hybrid explainability approach for ER, {\uncerta}, which combines self- with post-hoc explanation methods to mitigate scalability limitations.
\item We provide three new synthetic datasets for the ER task, that current LLMs have not previously had access to.
\end{itemize}
\noindent
Source code and datasets are available in a public repository.\footnote{\url{https://github.com/tteofili/ellmer}}

\myparagraph{Outline} Section~\ref{sec:foundations} presents preliminary notions. Section~\ref{sec:methods} illustrates self-explanations, post-hoc and hybrid explanation methods. Section~\ref{sec:experiments} reports experimental results. Section~\ref{sec:related} discusses related work, and Section~\ref{sec:conclusions} concludes the paper. 

\section{Foundations}
\label{sec:foundations}

We now introduce the core concepts underlying our analysis. 

\smallskip
\myparagraph{Entity Resolution Problem} We consider the clean-clean entity resolution setting, where two data sources $U$ and $V$ are individually (almost) duplicate-free.
Formally, given two sets of records $U$ and $V$, the goal is to classify record pairs in $E=U \times V$. 
Given a pair of records $(u, v) \in U \times V$, we define $M(u,v) = y$ as the prediction generated by a model $M$; $y \in \{{T}, {F}\}$, 
where ${T}$ indicates a \emph{match}, that is, $u$ and $v$ refer to the same entity, whereas ${F}$ indicates a \emph{non-match}, that is, $u$ and $v$ refer to different entities. 

\smallskip
\myparagraph{Solving ER with LLMs} We focus on models that are LLMs and thus we interact with them via prompting. In the prompt, we represent records as sequences of tokens $u=t_1, t_2, \dots, t_k$ where the schema can be loosely specified, such as $u=$``Model: Canon EOS 5D Mark IV, Resolution: 30.4 megapixels, Sensor: Full-frame CMOS''. 
For sake of simplicity, we use the symbol $u$ both to refer to the record and the corresponding set of tokens $\{t_1, t_2, \dots\, t_k\}$. In the prompt, we also add instructions on the ER task and on how to solve it, requiring the answer ``yes'' for a matching prediction and ``no'' for a non-matching prediction. 
Further details depend on the specific prompting strategy, which are detailed in Section~\ref{sec:selfexpl}.

\smallskip
\myparagraph{Explanations for ER} We consider \emph{feature attribution} and \emph{counterfactual} explanations, which are widely adopted for the ER task and can be evaluated by means of established metrics in the explainable AI literature~\cite{saliency_metrics,cf_metrics}. 

More specifically, in our context:
\begin{itemize}[leftmargin=*]
    \item 
    a \emph{feature attribution} explanation assigns an \emph{attribution score} to each feature in $u \cup v$, capturing its contribution to the  outcome of $M(u, v)$;%
    \item
    a \emph{counterfactual} explanation provides instructions on how to change features in $u \cup v$ in order to flip the outcome of $M(u, v)$.
\end{itemize}

Technically, a feature attribution explanation consists of a map that associates each feature with a numerical score, while a counterfactual explanation consists of a pair $(u^{\prime}, v^{\prime})$ that is equal to $(u, v)$ except for one or few feature values. 

Throughout this work we use the term \emph{feature} as a unifying abstraction to refer to either of two granularities at which explanations can be produced: \emph{attribute} or \emph{token}. Following standard database terminology, an \emph{attribute} denotes the complete value of a single field in a record (e.g., the entire \texttt{title}
or \texttt{price} value), whereas a \emph{token} denotes an individual word within an attribute value. Attribute-level explanations assign a single feature attribution score to each field and propose counterfactual modifications at the field level; token-level explanations operate at finer granularity, scoring and modifying individual words. The two granularities are not mutually exclusive, as they address different user needs~\cite{lipton2018mythos, guidotti2018survey}.

To give an example, consider the records $u=$``Model: Nikon D850, Sensor: Full-frame CMOS'', and $v=$``Model: Nikon D750, Sensor: CMOS'', with $M(u, v) = {F}$. A token-level feature attribution explanation might assume the form \{D850$\rightarrow$.8, D750$\rightarrow$.6, Full-frame$\rightarrow$.1, \dots\}, highlighting the importance of tokens such as ``D850'' and ``D750''. Similarly, an attribute-level feature attribution explanation might be \{Model: Nikon D750$\rightarrow$.7, Model: Nikon D850$\rightarrow$.6, Sensor: Full-frame CMOS$\rightarrow$.2\, Sensor: CMOS$\rightarrow$.01\}, revealing that the value of the Model attribute in $u$ is the most relevant for the prediction. A counterfactual token-level explanation could indicate that changing  the token ``D850'' in $u$ into ``D750'' would flip the prediction from ${F}$ to ${T}$. 

\smallskip
\myparagraph{Faithful and actionable explanations.}
An explanation is \emph{faithful} if it accurately reflects the model's actual decision process, as opposed to merely being plausible or coherent to a data steward or analyst reviewing the output~\cite{jacovi2020towards,lipton2018mythos}. This distinction is relevant because a fluent, human-readable explanation can
be compelling without necessarily corresponding to the features the model actually relied upon. For feature attribution explanations, faithfulness is operationalised through feature
removal tests: a faithful ranking tends to assign high scores to features whose masking causes the largest degradation in model performance~\cite{samek2016evaluating,deyoung2020eraser}. Importantly, faithfulness is defined relative to the \emph{model's} decision
process rather than to ground truth, so a faithful explanation for a wrong prediction is not necessarily a contradiction: it may accurately characterise \emph{why the model erred}, which can be useful for auditing, error analysis, and targeted correction. This connection to \emph{actionability}~\cite{wachter2017counterfactual,karimi2021algorithmic} gives faithfulness practical relevance beyond its technical definition. We provide experimental evidence of this in Section~\ref{sec:correctness}.

\section{Self-Explanation and Post-Hoc Methods}
\label{sec:methods}

This section introduces the explainable AI techniques for ER considered in our analysis. We distinguish between two main families of approaches: \emph{self-explanations} and \emph{post-hoc explanations}. Both aim at providing insight into ER predictions through feature attribution and counterfactual explanations. Then, we introduce a \emph{hybrid} approach, which combines self- with post-hoc explanations. 

Self-explanation approaches rely on large language models that are prompted to jointly produce predictions and explanations within the same model invocation, explicitly clarifying the rationale underlying their decisions. This paradigm, introduced in~\cite{llm_selfexplain}, is discussed in Section~\ref{sec:selfexpl}, where we describe how different prompting strategies are tailored to the ER task.

Post-hoc approaches, instead, generate explanations after predictions have been produced, by analyzing the behavior of the model through targeted perturbations of input features and observing the resulting changes in its outputs~\cite{slack2021reliable,madsen2022post}. Section~\ref{sec:posthoc}  illustrates the SOTA post-hoc explanation methods adopted in our analysis.

The hybrid approach aims to leverage the strengths of both self- and post-hoc explanations, and it is presented in Section~\ref{sec:hybrid}. 

\subsection{Self-Explanations}
\label{sec:selfexpl}

Thanks to the flexibility granted by prompting ~\cite{liu2023pre,brown2020language}, LLMs can be instructed to return different forms of self-explanations. We are interested in obtaining structured responses  representing feature attribution and counterfactual explanations.
Based on previous work~\cite{nananukul2024cost,wei2022chain,brown2020language}, we leverage three different prompting approaches: \emph{Zero-Shot}, \emph{In-Context Learning}, \emph{Chain-of-Thought}.

\myparagraph{Feature attribution explanations.} Across all three prompting strategies, the LLM is asked to assign a numerical score to each input feature, at either attribute or token granularity, reflecting how much that feature is perceived to contribute to the matching decision. The prompt specifies the output format (a score per feature, returned as a structured JSON object) and provides a natural-language definition of what the score represents, but imposes no constraint on how scores should be derived.

\myparagraph{Counterfactual explanations.} For counterfactuals, the model is asked to produce a modified version of the input pair such that the predicted label would flip: a non-match pair becomes a match, or vice versa. As with feature attribution explanations, the prompt defines the target output structure (a record pair with one or more feature values altered) and leaves the construction strategy to the model. The expected behavior (minimal and semantically coherent edits that are sufficient to reverse the prediction) is implied by the task description.

After initialization with a short description of the ER task, the three approaches work as follows. 

\begin{figure}[t]
    \centering
    \includegraphics[width=0.95\columnwidth]{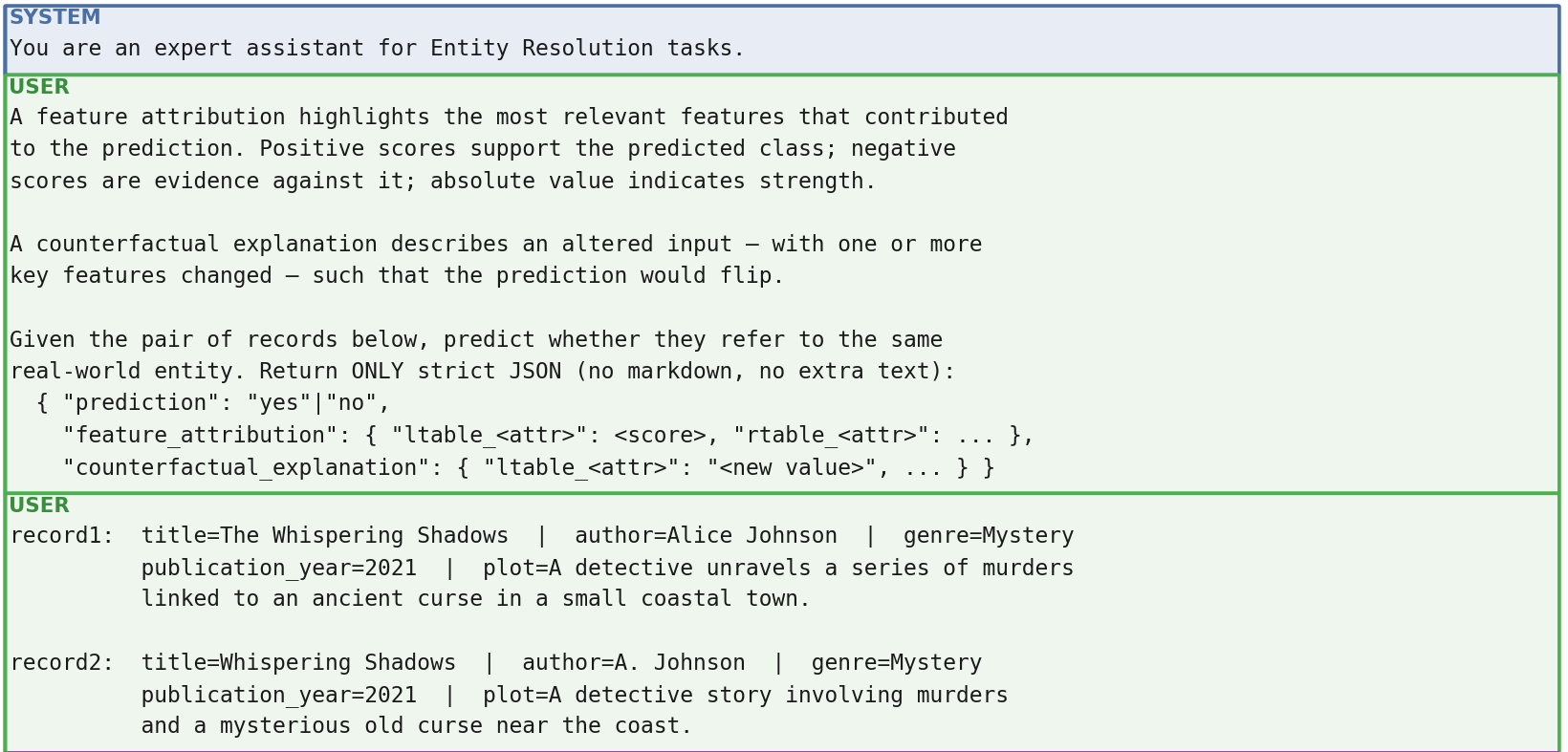}
    \caption{Example of a Zero-Shot prompt.}
    \Description{Example of a Zero-Shot prompt.}
    \label{fig:zs}
\end{figure}

\myparagraph{Zero-Shot (ZS)} As described in ~\cite{li-2023-practical}, we rely on the simple and cost-effective zero-shot prompt, where no prior context is given to the LLM about other records from the dataset. The only context we provide is a textual description of what feature attribution and counterfactual explanations consist of. 
Figure~\ref{fig:zs} shows an example of a ZS prompt. 
First, the prompt illustrates to the LLM what feature attribution and counterfactual explanations are. Then, it prescribes to the LLM the expected output as a JSON object with \emph{(i)} a yes/no response indicating the matching/non-matching prediction, \emph{(ii)} a feature attribution explanation table having a column for each input feature containing each feature’s score, \emph{(iii)} a counterfactual explanation table having a column for each input feature containing its counterfactual value. Finally it provides the pair of records ($record1$ and $record2$) to be processed.

\begin{figure}[t]
    \centering
    \includegraphics[width=0.95\columnwidth]{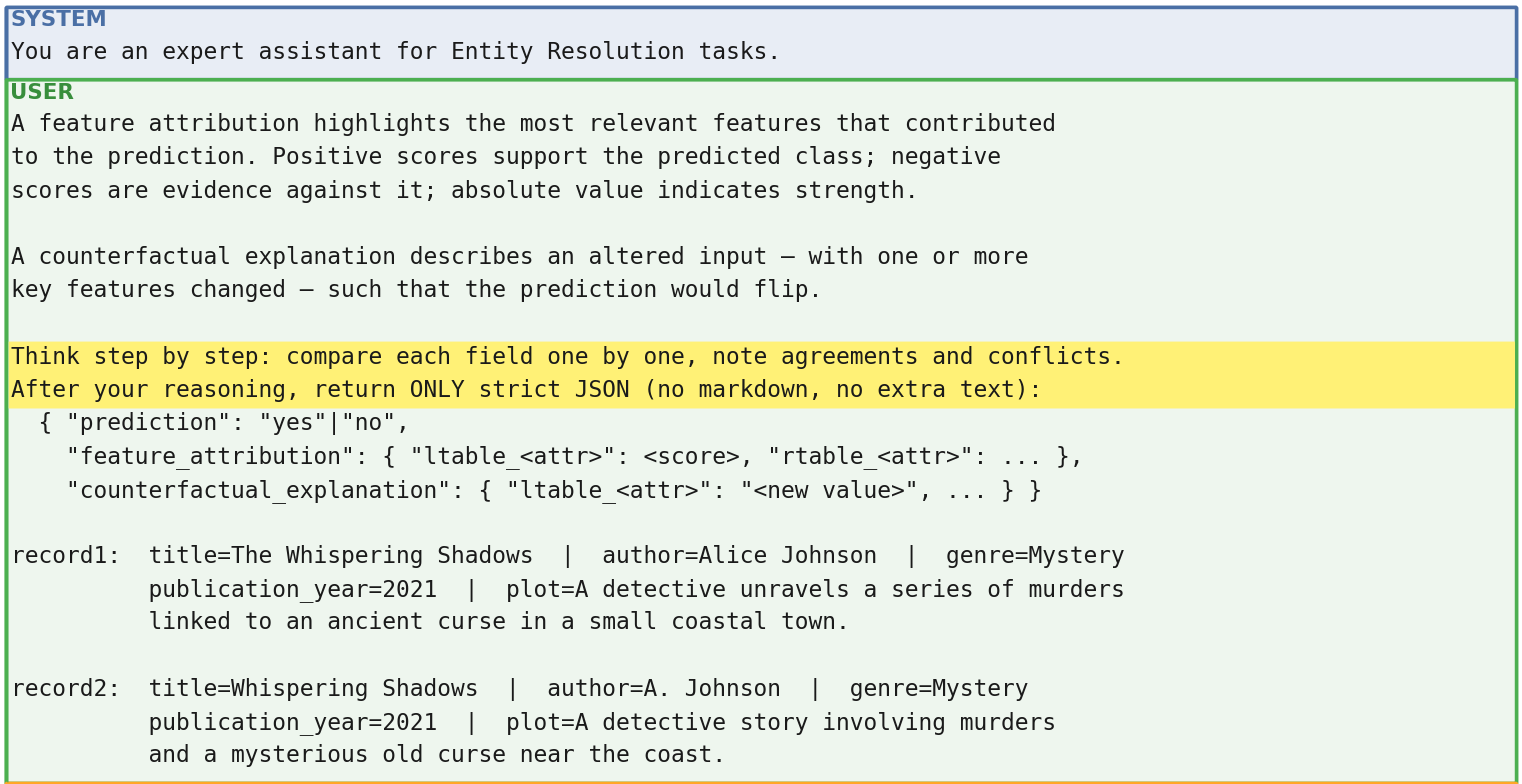}
\caption{Example of a Chain-of-Thought prompt.
}
\Description{Example of a Chain-of-Thought prompt}
    \label{fig:cot}
\end{figure}

\myparagraph{Chain-of-Thought (CoT)} Using the Chain-of-Thought prompting technique~\cite{wei2022chain}, we prompt the LLM to generate explicit intermediate reasoning steps leading to its final prediction for the entity resolution explanation task. In contrast to the zero-shot setting, where predictions and explanations are produced directly without an explicit reasoning structure, CoT encourages the model to decompose the decision process into successive steps, making the resulting explanations more structured and transparent.
Figure~\ref{fig:cot} provides an example of a CoT prompt.

\begin{figure}[t]
    \centering
    \includegraphics[width=0.95\columnwidth]{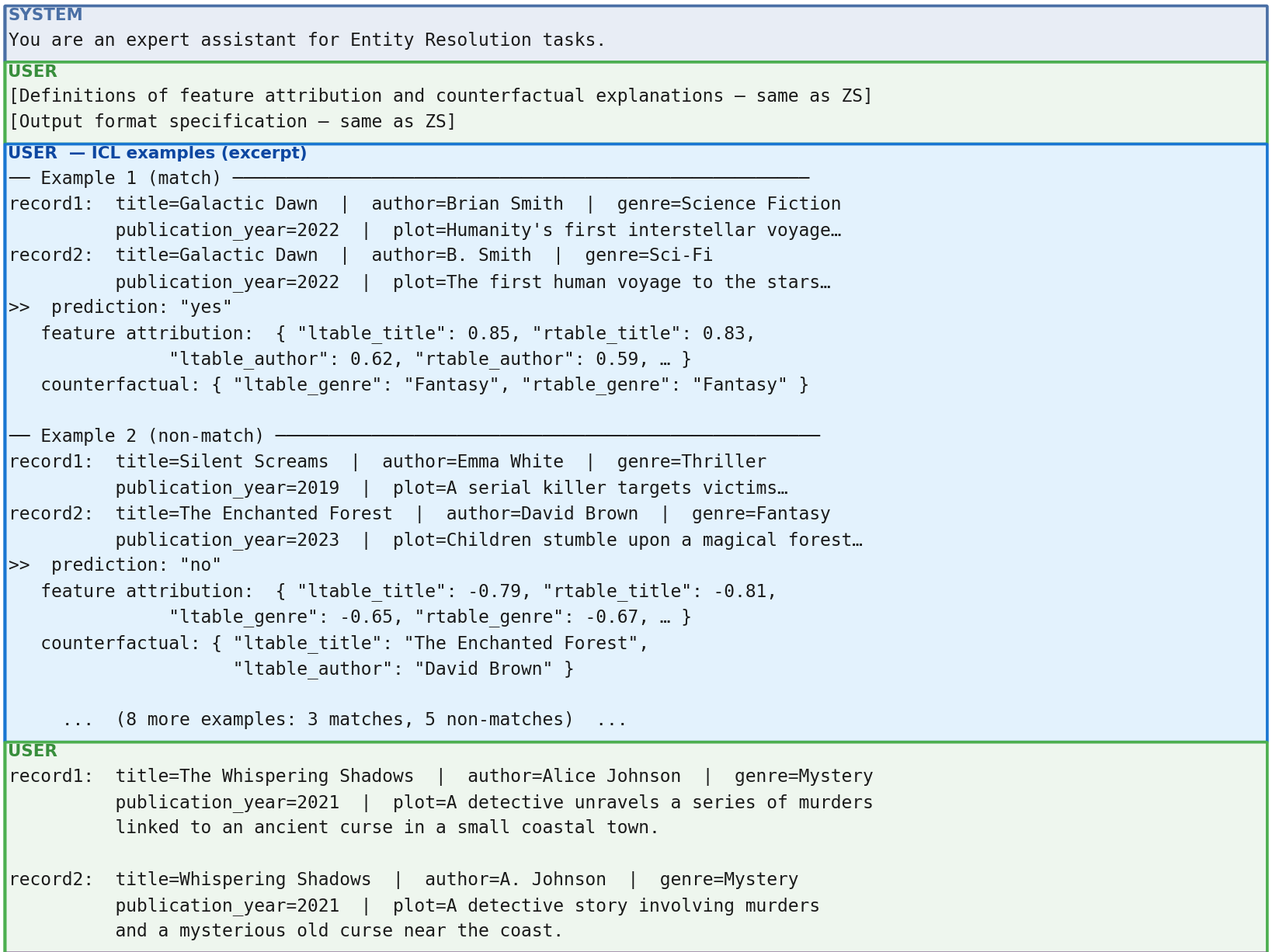}
    \caption{Example of an In-Context Learning prompt.}
    \Description{Example of an In-Context Learning prompt.}
    \label{fig:icl}
\end{figure}

\myparagraph{In-Context Learning (ICL)} Consistent with prior work showing that In-Context Learning can improve LLM performance on downstream tasks~\cite{brown2020language}, we augment the ZS prompt with a set of labeled examples. Specifically, we provide ten examples (five matching and five non-matching pairs) each accompanied by feature attribution and counterfactual explanations, allowing the model to condition its predictions and explanations on observed patterns.
Figure~\ref{fig:icl} reports a portion of the examples that we provide in an ICL prompt.

\subsubsection{Eliciting vs.\ Instructing Explanations.}
A key design choice in our evaluation of self-explainers is that we constrain the prompts to specify \emph{what kind} of explanation the model should produce, a ranked feature attribution over attributes or tokens, and a minimal counterfactual record pair, but deliberately leave unspecified \emph{how} that explanation should be computed. 
This choice reflects the nature of the object being evaluated: a self-explanation is not the output of an explicit algorithmic procedure invoked by the user, but rather what the LLM produces \emph{naturally} as a complement to its matching prediction.
Imposing a specific computational method in the prompt would shift the task from \emph{eliciting} an explanation to \emph{instructing} one, conflating the model's native reasoning with externally prescribed procedure.
As we show in Section~\ref{sec:introspection}, this distinction is meaningful in practice: when asked to describe the procedure they followed to assign feature attribution scores and construct counterfactuals, different models gave qualitatively different (and largely inconsistent) answers.

Instructing the model on \emph{how} to compute its explanations would change the object of study: we would be measuring the model's ability to execute a prescribed procedure, not the reliability of the explanations it produces autonomously as part of its inherent prediction workflow.

\subsection{Post-hoc explanations}
\label{sec:posthoc}
Post-hoc methods for the ER task include  Mojito~\cite{mojito}, Explainer~\cite{ ebaid2019explainer}, Landmark~\cite{baraldi2021landmark}, \minun{}~\cite{minun},  \certa{}~\cite{teofili2022effective}, and  \lemon{}~\cite{lemon}. 

Mojito, Landmark, and \lemon{} build upon the seminal LIME approach~\cite{ribeiro2016should}.

In our experimental analysis, we use \certa{}, \lemon{}, and \minun{} as post-hoc explanation methods as they represent among the most recent and reliable solutions specifically designed for the ER task, and have been evaluated on standard  benchmarks.\footnote{We extend the original implementation of \certa{}\cite{teofili2022effective}, which is limited to attribute-level explanations, to support explanations at both the attribute and token levels.}
\certa{} and \lemon{} generate feature attribution explanations, \certa{} and \minun{} provide counterfactual explanations, enabling a direct comparison with self-explanations when addressing the research questions on trustworthiness, effectiveness, and efficiency.
 
\lemon{}~\cite{lemon} generates perturbed variants of the input pair $(u,v)$ by masking or replacing individual features, queries $M$ on each variant, and fits a weighted linear surrogate model $g$ on the resulting (perturbation, prediction) pairs using a LIME-style proximity kernel. 
The attribution score of each feature is the corresponding coefficient in $g$, reflecting how much that feature's presence influences the local linear approximation of $M$'s decision boundary around $(u,v)$.

\certa{}~\cite{teofili2022effective} produces both feature attribution and counterfactual explanations via a probabilistic perturbation framework.
\certa{} builds on the concept of \emph{open triangles} over $(u, v)$ by finding another record $w \in U$ (called \emph{support record}) such that the ER system predicts $(w,v)$ to have a different prediction with respect to $M(u,v)$.
\emph{Perturbing} $u$ by progressively copying attribute values from $w$ to $u$, deriving a $u'$, increasingly making $u'$ more similar to $w$ based on their content, at some point the prediction of the model will flip, declaring $u'$ and $v$ such that $M(u,v) \neq M(u',v)$.
Repeating the same procedure for many support records produces evidence of the influence that attributes and sets of attributes have on the input prediction. 
For feature attribution explanations, the attribution score of an attribute is defined as the probability that changing the value of that attribute is a \emph{necessary} factor for flipping the outcome of the prediction (\emph{probability of necessity}), across different open triangles. Symmetrically, \certa{} generates \emph{counterfactual} explanations having the highest probability that changing the value of a certain \emph{set of attributes} is a \emph{sufficient} factor for flipping the outcome of a prediction (\emph{probability of sufficiency}).
Such probabilities are computed by a frequentist approach. The number of times an attribute is changed over the number of actual flips gives the \emph{probability of necessity}, the number of times changing a set of attributes results in a flip over the number of times that the set of attributes is changed gives the \emph{probability of sufficiency}.
\\
\minun{}~\cite{minun} is a post-hoc method for the ER task aimed at generating  counterfactual explanations. It adopts an approach based on perturbations, which are produced by modifying the input pair until the original prediction flips. The perturbation is performed by 
considering all possible sequences of token insertions, removals, and replacements to transform one record (e.g., $u$ in $(u, v)$) into the other. Greedy and approximate algorithms efficiently compute the shortest transformations.

\subsection{{\uncerta}: a Hybrid Explanation Method}
\label{sec:hybrid}

Post-hoc explanations require performing a large number of predictions over the perturbed samples. In the LLM scenario, this yields significant computational (and monetary) cost because each prediction corresponds to a call to a LLM. On the other hand, self-explanation methods can generate explanations at the cost of a single prediction but their truthfulness is not guaranteed. Although self-explanations can highlight seemingly reasonable features, there is no guarantee that these are the features that truly influenced the prediction by the LLM. 

For these reasons, we introduce a hybrid method, {\uncerta}, which leverages the strengths of both self- and post-hoc explanations. 

The main intuition of {\uncerta} is to use self-explanations as a prior to optimize search for flip-producing feature changes in the perturbation space. Reducing the number of explored perturbations can directly reduce the number of predictions (and thus the computational cost) without compromising on trustworthiness. 

Given a prediction $M(u, v)=y$ produced by an LLM $M$, let $P = \{p_1, \dots, p_n \}$, be a set of prompts used for generating  self-explanations. 
For each $p \in P$, we collect the top $k$ features of feature attribution self-explanations into a list $\Xi$. 
In our implementation, $P$ includes the three prompts (ZS, ICL, CoT) described in Section~\ref{sec:selfexpl}.
We empirically observed that adding other prompts (e.g., using different forms of CoT and examples in ZS) does not produce significant improvements.

In the worst case, $\Xi$ may contain up to $n \cdot k$ features, when each of the $n$ self-explanations yields a disjoint set of top-$k$ features. To bound the size of $\Xi$, we apply a frequency-based selection strategy: we count how often each feature appears across the $n$ self-explanations and retain only the $k$ most frequent ones, thereby favoring agreement across different prompts.

We call $\Xi^*$ the final set of features (attributes or tokens) selected from $\Xi$. This set is used as a \emph{perturbation mask} to be passed to post-hoc algorithms. 
Note that any post-hoc algorithm can be used to generate a feature attribution explanation, a counterfactual explanation or both, provided that it can be made to focus only on the features specified in $\Xi^*$, to reduce the search space.

To avoid cases in which the total attribution scores of the selected feature set $\Xi^*$ is negligible (indicating the absence of sufficiently informative features), we require the sum of attribution scores over $\Xi^*$ to exceed a threshold $\phi$. 
If this condition is not satisfied, we iteratively increase $k$ until either the threshold $\phi$ is met or $\Xi^*$ expands to include all available features.
The values of $\phi$ and $k$ are determined through empirical evaluation. In our experiments, we set $\phi = 0.5$, $k = \frac{|\textnormal{features}|}{2}$ for attribute-level explanations, and $k = 5$ for token-level explanations, as these values provided a good trade-off between coverage and compactness.

In our experimental evaluation, we instantiate \uncerta{} with different post-hoc explanation methods depending on the explanation type. For feature attribution explanations, we employ \lemon{} and \certa{}, while for counterfactual explanations we use \minun{} and \certa{}. Throughout the paper, we denote these instantiations as \uncerta{}$_L$, \uncerta{}$_C$, and \uncerta{}$_M$, corresponding to hybrid explanations built upon \lemon{}, \certa{}, and \minun{}, respectively. For the attribute-level explanations, we populate $\Xi^*$ with attribute values (e.g., ``Nikon D750''), whereas for the token-level version we fill $\Xi^*$ with individual tokens (e.g., ``Nikon''). 

\begin{table*}[th]
\caption{Datasets details. For each dataset, we report: sizes of left and right tables ($|L|$ and $|R|$), total number of candidate pairs ($|L \times R|$), test set size($|test|$), number of attributes ($\textit{attr.}$), average number of tokens per record ($\textit{tokens}$), F1 score of ER predictions obtained with the three LLMs.} 
\label{fig:datasets}
\renewcommand{\arraystretch}{.78}
    \centering
    \small
    \begin{tabular}{ccccccccccc}
\multicolumn{1}{c}{} & &  &  &  &  & & & \multicolumn{3}{c}{F1-score} \\
\cmidrule{9-11}
\multicolumn{1}{c}{} & Dataset & |L| & |R| & $|L \times R|$ & |test| & attr. & tokens & LlaMa3.1-8B & ChatGPT-5 & Claude-4.6 \\
\toprule
DeepMatcher & AB & 1,081 & 1,092 & 5,743 & 1,916 & 3 & 59 &  0.93 & 0.93 & 0.91\\
& BR & 4,345 & 3,000 & 68 & 91 & 4 & 26 & 0.90 & 0.95 & 0.94\\
& FZ & 533 & 331 & 110 & 189 & 6 & 37  & 1 & 1 & 1\\
& WA & 2,554 & 22,074 & 962 & 2,049 & 5 & 15 & 0.90 & 0.93 & 0.80\\
& AG & 1,363 & 3,226 & 1,167 & 2,293 & 3 & 10 & 0.88 & 0.95 & 0.82 \\
\midrule
WDC & Cameras & 17,128 & 17,128 & 16,028 & 1,100 & 7 & 229  & 0.62 & 0.97 & 0.95 \\
& Watches & 22,721 & 22,721  & 21,621 & 1,100 & 7 & 193 &  0.59 & 0.96 & 0.95\\
\midrule
Synthetic & FB & 520 & 495 & 100 & 317 & 5 & 35 & 0.77 & 0.81 & 0.85\\
& FCP & 500 & 600 & 220 & 163 & 8 & 37 & 0.74 & 0.79 & 0.80 \\
& FakER & 300 & 300 & 150 & 350 & 210 & 776 & 0.83 & 0.99 & 0.93\\
\bottomrule
\end{tabular}
\end{table*}

\section{Experiments}
\label{sec:experiments}

In this section we discuss experimental results related to the research questions introduced in Section~\ref{sec:intro}.

\myparagraph{LLMs and Prompting}
We consider three representative large language models: Claude-4.6-Sonnet~\cite{anthropic_claude_4_6_sonnet} (April 2026), ChatGPT-5.0 (December 2025), and LLaMA~3.1-8B (December 2025). These models were selected to capture complementary design choices in terms of architecture, scale, training regimes, and deployment settings, including both closed-source and open-weight models. This selection allows us to assess whether the observed behavior of self-explanations is consistent across heterogeneous LLM configurations rather than being specific to a single model.
To ensure reproducibility of results, we set the temperature parameter to $0$ for all models in all reported experiments. This reduces stochastic variations in the LLMs, making the results comparable across runs.
For each LLM, self-explanations are generated using the prompting strategies described in Section~\ref{sec:selfexpl}, enabling a systematic comparison of the robustness of explanations under different prompting conditions. Whenever possible, we leverage \emph{optimized batch prompting}~\cite{ji2025optimized}, to mitigate the inherent latency amplification in LLMs inference.

\myparagraph{Datasets}
We evaluate entity resolution predictions produced by the considered LLMs on selected datasets from the DeepMatcher~\cite{mudgal2018deep} and WDC~\cite{primpeli2019wdc} benchmarks, which are widely used for semantic entity resolution. In particular, we consider the Abt-Buy (AB), BeerAdvo-RateBeer (BR), Fodors-Zagats (FZ), Walmart-Amazon (WA), and Amazon-Google (AG) datasets from DeepMatcher,\footnote{\url{https://github.com/anhaidgroup/deepmatcher/blob/master/Datasets.md}} and the Cameras and Watches datasets from WDC.\footnote{\url{https://huggingface.co/datasets/wdc/products-2017}} 

Since these datasets may have been observed during LLM pre-training, we additionally introduce two novel synthetic datasets, consisting of descriptions of \emph{fictional books} (FB) and \emph{fictional car parts} (FCP), generated using an independent LLM (Mistral Nemo~\cite{jiang2023mistral}). Each dataset consists of two table sources $L$ and $R$, a labeled training set $G$, and a labeled test set $T$, with an overall target size $s$. Initially, both sources are empty, i.e., $L = \emptyset$ and $R = \emptyset$. Then, the datasets are generated according to the following procedure:

\begin{enumerate}[leftmargin=*]
\item For a given domain, we use Mistral to generate two disjoint sets of entity descriptions, $A$ and $B$, with $A \cap B = \emptyset$.

\item From $A$, we randomly select one third of the entities and generate alternative textual representations that preserve the same underlying identity, obtaining a set $C$.

\item Positive (matching) pairs are generated by sampling $\langle a, c \rangle$, with $a \in A$ and $c \in C$, and are evenly split between the training set $G$ and the test set $T$ (label $=1$).

\item Negative (non-matching) pairs are generated by sampling $\langle a, b \rangle$, with $a \in A$ and $b \in B$, and are evenly split between $G$ and $T$ (label $=0$).

\item The two sources are updated as $L = L \cup A$ and $R = R \cup B \cup C$.

\item The procedure is repeated until $|L| + |R| $ reaches $s$.
\end{enumerate}

For the FB dataset, we additionally generate several batches of samples in languages other than English (Spanish, German, and Italian) to increase linguistic diversity and overall dataset difficulty.

We also construct a synthetic ER dataset (dubbed \emph{FakER}), designed as a worst-case scenario to assess scalability and robustness (rather than realism), to evaluate explanations generated under high attribute dimensionality. 
Each source contains $300$ records, comprising $150$ shared entities appearing in both sources and $150$ source-specific entities unique to each source. Shared entities form the positive matches across sources, while source-specific entities act as natural distractors. All records follow a common schema with $105$ textual attributes and a unique source-specific identifier.
Attribute values are entirely textual and generated using a random mixture of realistic data types, including personal names, postal addresses, company names, short natural-language sentences, and short free-text fragments. This design reflects ER scenarios in which entities are described by large numbers of heterogeneous textual fields, rather than a small set of curated attributes.
To simulate real-world data inconsistencies, we introduce controlled textual variations independently in each source. Starting from a clean base representation for each shared entity, we apply attribute-level noise including character-level typos, abbreviations and truncations, synonym substitutions (e.g., street -> st), missing or extra characters, and limited word reordering. The noise level is asymmetric across sources: medium variation in Source A and high variation in Source B, reflecting differences in data quality commonly observed in practice. All random processes are seeded to ensure reproducibility.
Ground truth consists of $150$ positive record pairs corresponding to shared entities and $200$ randomly sampled negative pairs drawn from non-matching record combinations.

For each dataset, the LLMs are prompted on all record pairs in the test set to predict match/non-match labels.
Table~\ref{fig:datasets} reports the main features of the datasets and the F-1 measure of the predictions returned by the LLMs that we have used in the experiments.

\myparagraph{Metrics}
The evaluation of explainability results aims to verify that the explanations accurately capture the features that truly influenced the prediction~\cite{jeyakumar2020can}. 
We use two popular trustworthiness metrics from the explainable AI literature~\cite{coroama2022evaluation}: \emph{faithfulness}~\cite{saliency_metrics} for feature attribution,  \emph{validity}~\cite{cf_metrics} for counterfactual explanations. 

The main idea behind the \emph{faithfulness} metric is that modifying features with high attribution score should cause a significant change in the model predictions, whereas modifying features with low scores should have a negligible effect. For each dataset, we consider all record pairs and progressively remove, for each pair, a fraction $\theta$ of the \emph{most important} features (attributes or tokens), according to their attribution scores. For instance, $\theta = 0.3$ means that the top $30\%$ of features in the non-increasing attribution score ranking are removed. We then compute a curve that maps $\theta$ to the resulting prediction performance, measured in terms of F1-score. Faithfulness is defined as the area under this curve~\cite{saliency_metrics}. Highly trustworthy explanations are expected to yield a smaller area, as removing even a small fraction of highly important features should lead to a sharp degradation in prediction performance. In our experiments, we set $\theta \in \{0.1, 0.2, 0.33, 0.5, 0.7, 0.9\}$.

For counterfactual explanations, we assess their quality using the \emph{validity} metric, which measures the proportion of proper counterfactuals. Specifically, for each dataset, we apply the changes prescribed by the counterfactual explanations to the corresponding record pairs and re-query the LLM. Validity is then computed as the fraction of cases in which the model’s prediction is successfully flipped, either from match to non-match or vice versa.

All experiments are repeated five times and results are reported as means. Run-to-run variability is low (coefficient of variation <5\% across all metrics and methods). Statistical significance is assessed using paired two-sided t-tests across datasets, comparing the mean performance over the five runs for each method. Holm–Bonferroni correction~\cite{holm1979simple} is applied to control the family-wise error rate, with significance assessed at p<0.05.

\subsection{RQ-1: Are self-explanations trustworthy?} 

To address RQ-1, we analyze the trustworthiness of LLM-generated self-explanations for ER along two complementary dimensions. First, we investigate the sensitivity of self-explanations to different prompting strategies, assessing whether variations in prompts lead to unstable or inconsistent explanatory outputs. Second, we examine the extent to which self-explanations are affected by hallucinations and evaluate their impact on the reliability of explanations in the ER setting. Finally, we analyze the consistency between attribute and token-level explanations, and the agreement between feature attribution and counterfactual explanations.  

\subsubsection{Sensitivity to Prompt Strategies}
\label{sec:propmt_sensitivity}
To assess prompt sensitivity in self-explanations, we first generate predictions and corresponding self-explanations for all record pairs considered in each dataset using the ZS, CoT, and ICL prompts. Then, for each prediction we measure the pair-wise similarity of the result of the three self-explanations strategies. Robust self-explanations should remain consistent even with different prompts.

\begin{figure}[t]
\subfigure[\normalsize{Avg.  Kendall-Tau correlation for feature attribution self-explanations ($\uparrow$).}]{
\includegraphics[width=\columnwidth]{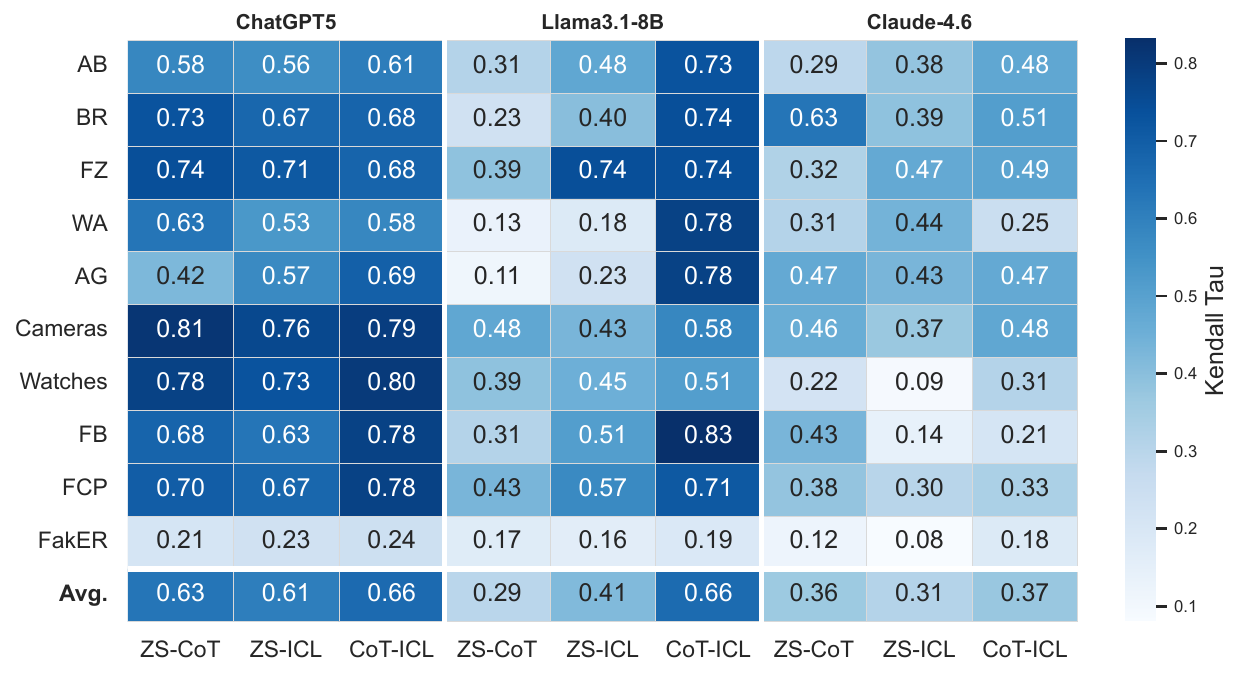}
}
\subfigure[\normalsize{Avg.  cosine similarity for counterfactual self-explanations ($\uparrow$).}]{
\includegraphics[width=\columnwidth]{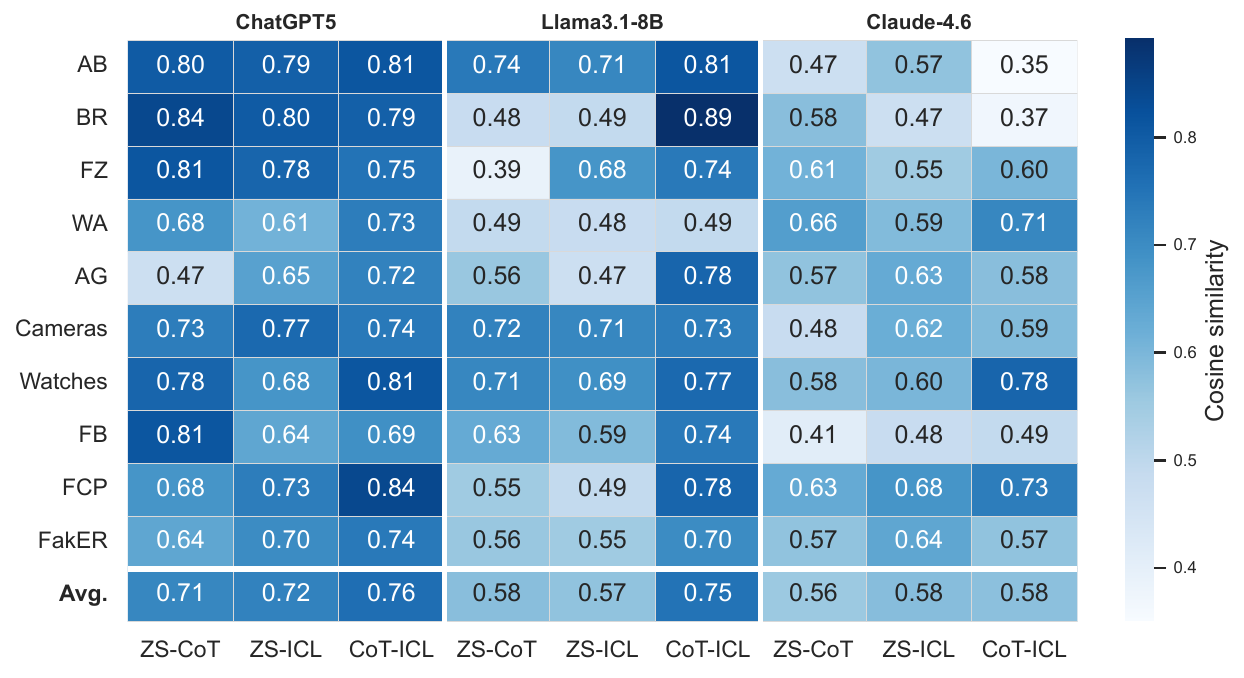}
}
\caption{Agreement of self-explanations (attribute-level); 1 (darkest) corresponds to the highest agreement ($\uparrow$).}
\Description{Agreement of self-explanations (attribute-level); 1 (darkest) corresponds to the highest agreement ($\uparrow$).}
    \label{tab:robustness}
\end{figure}

We evaluate the similarity of explanations as follows. For feature attribution explanations, we consider the relative importance attributed by the LLM to each feature, i.e., attribute or token, because absolute scores are more likely to change. Specifically, for each self-explanation we consider the ranking of features by non-increasing attribution score, and measure the Kendall-Tau correlation coefficient~\cite{kendall1948rank} of rankings corresponding to different prompts. A Kendall-Tau correlation of $1$ denotes complete agreement, whereas $-1$ indicates complete disagreement. A value of $0$ corresponds to no correlation. 
For counterfactual self-explanations we consider instead textual similarity. First, we transform each self-explanation into a bag-of-words vector weighted with TF-IDF score. Then we measure the cosine similarity of vectors corresponding to different prompts for the same prediction. 

For each dataset, Figure~\ref{tab:robustness} reports the average similarity  scores of \emph{ZS} self-explanations vs. \emph{CoT}, then \emph{ZS} vs. \emph{ICL}, and finally \emph{CoT} vs. \emph{ICL}. We report only attribute-level results as we observed that token-level results are very similar. In general, feature attribution explanations show suboptimal agreement, with Kendall-Tau often between $0$ and $0.5$, for Llama-3.1-8B and Claude-4.6-Sonnet. ChatGPT-5 shows a higher degree of agreement, across pairs, but still averaging a mild $0.6$ KT. For ChatGPT-5 and Llama-3.1-8B, \emph{CoT} and \emph{ICL} seem to agree more steadily. Claude-4.6-Sonnet shows a lower degree of agreement for feature attribution explanations.
Counterfactual explanations report high values of cosine similarity in some cases but also present high variability, suggesting that also counterfactual self-explanations are often in disagreement for the same prediction, when generated with different prompts. Also in this case we observe a higher degree of similarity between counterfactual explanations generated with \emph{CoT} and \emph{ICL} prompts for ChatGPT-5 and Llama-3.1-8B, whereas such similarity is generally lower for Claude-4.6-Sonnet.

\subsubsection{Faithfulness and Validity of Self-Explanations}
We now provide results for a key desiderata for self-explanations, that is, their ability to capture the actual matching criteria used by the LLM. 

\begin{figure}[t]
\centering
\subfigure[\normalsize{Attribute-level ($\downarrow$).}]{
\includegraphics[width=\columnwidth]{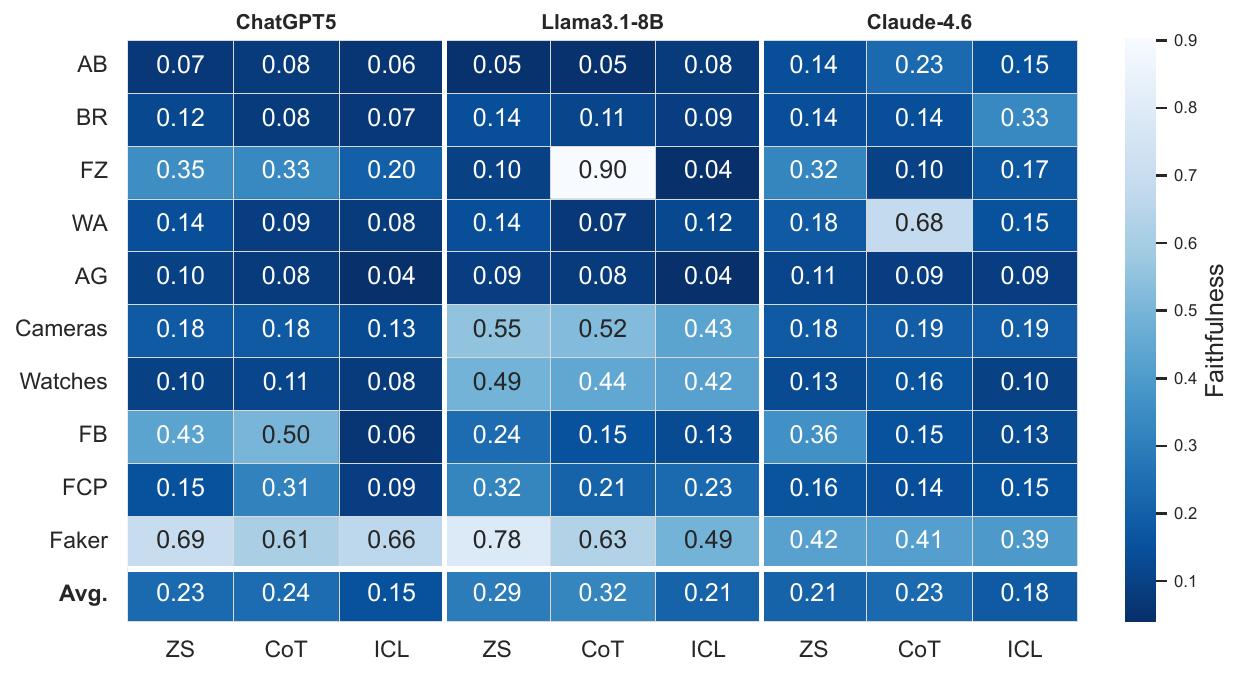}
}
\subfigure[\normalsize{Token-level ($\downarrow$).}]{
\includegraphics[width=\columnwidth]{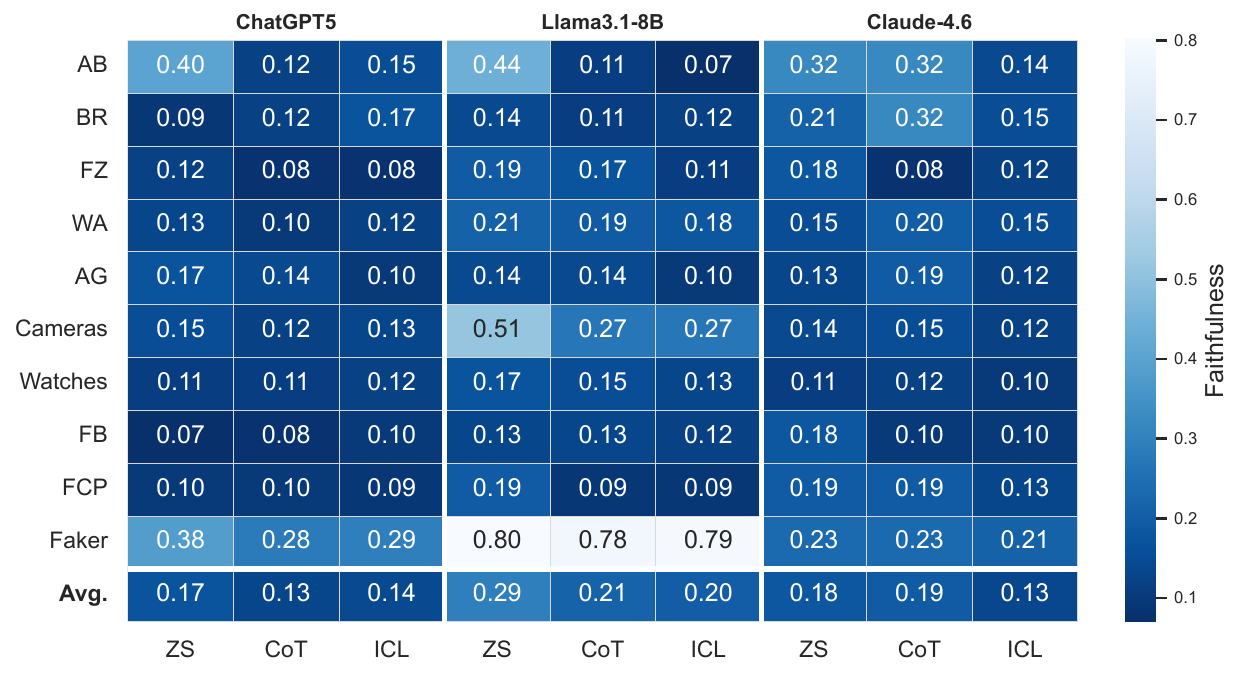}
}
\caption{Faithfulness of feature attribution self-explanations. Values close to 0 (darker) are better ($\downarrow$).}
\Description{Faithfulness of feature attribution self-explanations. Values close to 0 are better ($\downarrow$).}    
\label{fig:self-explanation-Faithfulness}
\end{figure}

\begin{figure}[ht]
\subfigure[\normalsize{Attribute-level ($\uparrow$).}]{
\includegraphics[width=\columnwidth]{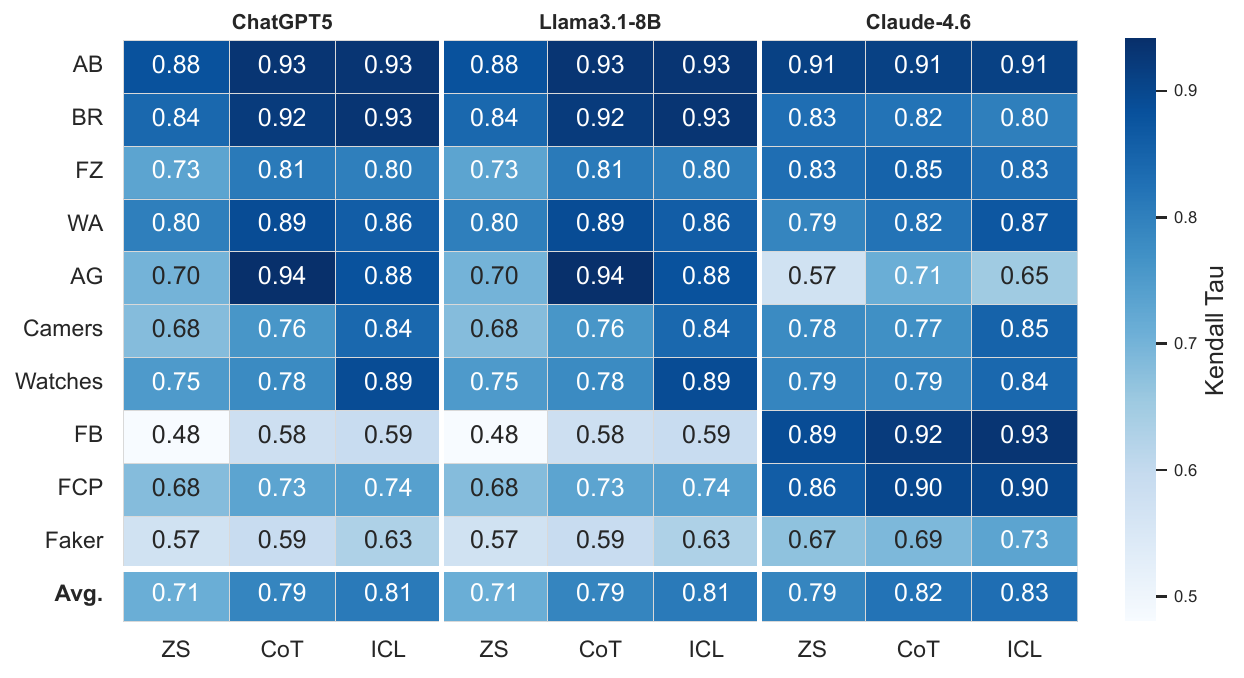}
}
\subfigure[\normalsize{Token-level ($\uparrow$).}]{
\includegraphics[width=\columnwidth]{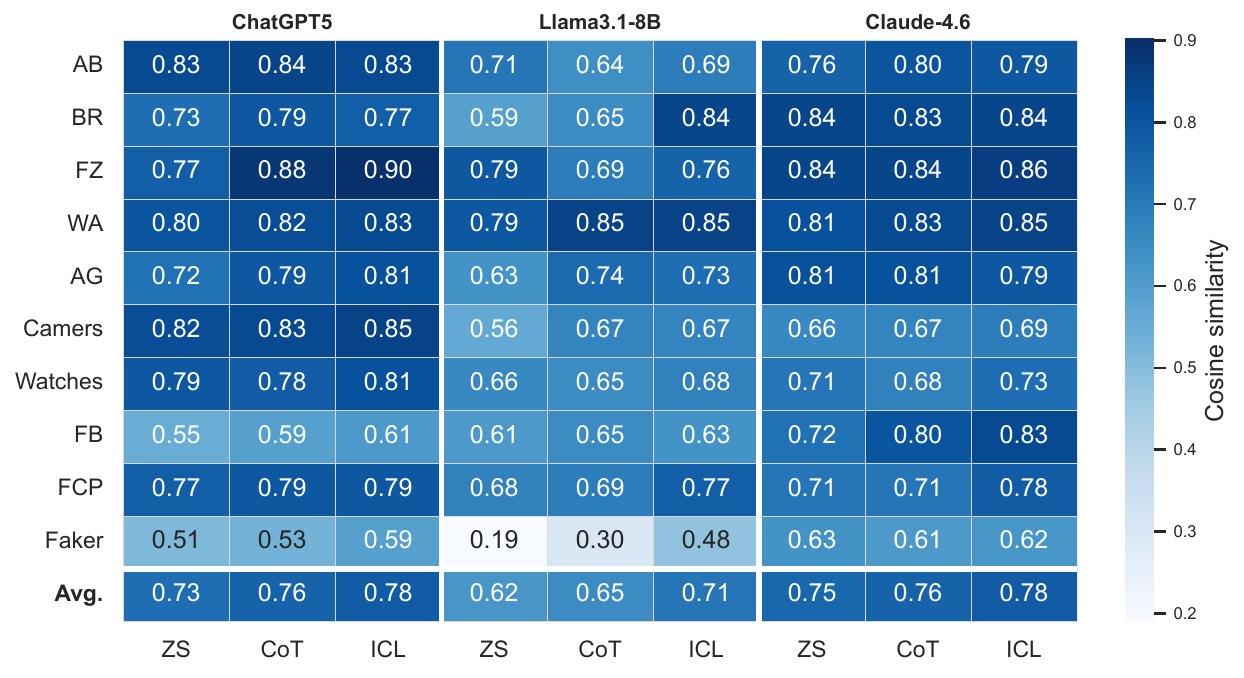}
}
\caption{Validity of counterfactual self-explanations. Values close to 1 (darker) are better ($\uparrow$).}
\Description{Validity of counterfactual self-explanations: (a) at attribute-level, (b) at token-level. Values close to 1 are better ($\uparrow$).}
    \label{fig:self-explanation-Counterfactual}
\end{figure}

Figure~\ref{fig:self-explanation-Faithfulness} reports the faithfulness results for all the datasets considered in our experiments at attribute and token level. 
With respect to prompting strategies, ICL consistently achieves better faithfulness than the other approaches across models, for both attribute-level and token-level explanations. An exception is observed for ChatGPT-5 at the token level, where CoT slightly outperforms ICL.
Similar observations hold for counterfactual self-explanations, whose results are presented in Figure~\ref{fig:self-explanation-Counterfactual}, but in this case the ICL strategy outperforms the others with all the models. 

ChatGPT-5 consistently produces the most trustworthy feature attribution explanations, at both the attribute and token levels. In contrast, Claude-4.6-Sonnet produces counterfactual explanations with a higher validity; while generally worse, LLaMA3.1-8B still exhibits comparable performance for both feature attribution and counterfactual explanations.

Overall, counterfactual explanations tend to be slightly more trustworthy at the attribute level than at the token level, although the observed differences strongly depend on the specific dataset and the LLM considered. 

Datasets characterized by a larger number of tokens (Cameras and Watches) do not display markedly different behaviors, suggesting that explanation quality does not scale proportionally with record length. All methods, however, struggle on the FakER dataset, where both faithfulness and validity substantially degrade.

The dominant pattern across models is that \textsc{ICL} and \textsc{CoT} tend to outperform \textsc{ZS}, yielding lower faithfulness and higher validity, but the statistical footprint of this ordering is both model and granularity-dependent.
The effect of prompting strategy is concentrated at the attribute level. For both ChatGPT-5 and LLaMA-3.1-8B, CoT and ICL significantly improve faithfulness and validity relative to zero-shot prompting, whereas differences largely disappear at the token level. The strongest gains are observed for ChatGPT-5, where ICL substantially outperforms zero-shot prompting on both metrics (all (p < 0.001)). However, the additional benefit of ICL over CoT is limited: the ICL--CoT comparison is never significant at the token level and becomes significant only for ChatGPT-5 at the attribute level.
Claude-4.6 exhibits a markedly different behavior. Across all twelve pairwise comparisons, only two reach statistical significance, both favouring ICL over zero-shot prompting: token-level faithfulness and attribute-level validity. All remaining comparisons are non-significant, indicating that self-explanation quality is largely insensitive to the choice of prompting strategy.
Overall, these results show that improved prompting can enhance self-explanation quality, particularly for ChatGPT-5 and LLaMA-3.1-8B at the attribute level. However, most of the benefit is already achieved with CoT prompting, leaving little room for further improvement through few-shot exemplars. For Claude-4.6, prompting strategy has only a marginal impact on explanation quality.

\subsubsection{Cross-Granularity Consistency}

While consistency across levels of granularity might be implicitly assumed, it is unclear whether token-level attributions reliably aggregate into meaningful attribute-level explanations. 
To investigate this aspect, we assess the consistency between attribute-level and token-level self-explanations. 
For feature attribution explanations, we compute the Kendall–Tau correlation between the ranking of attributes induced by their attribute-level scores and the ranking obtained by averaging the scores of their constituent tokens. A positive correlation indicates hierarchical consistency.
For counterfactual explanations, we evaluate cross-level consistency by computing the cosine similarity between the sets of attribute values identified as counterfactuals at the attribute level and the corresponding sets derived from token-level counterfactual explanations.

\begin{figure}[t]
\subfigure[\normalsize{Kendall-Tau correlation between attribute and token-level self-explanations ($\uparrow$).}]{
\includegraphics[width=\columnwidth]{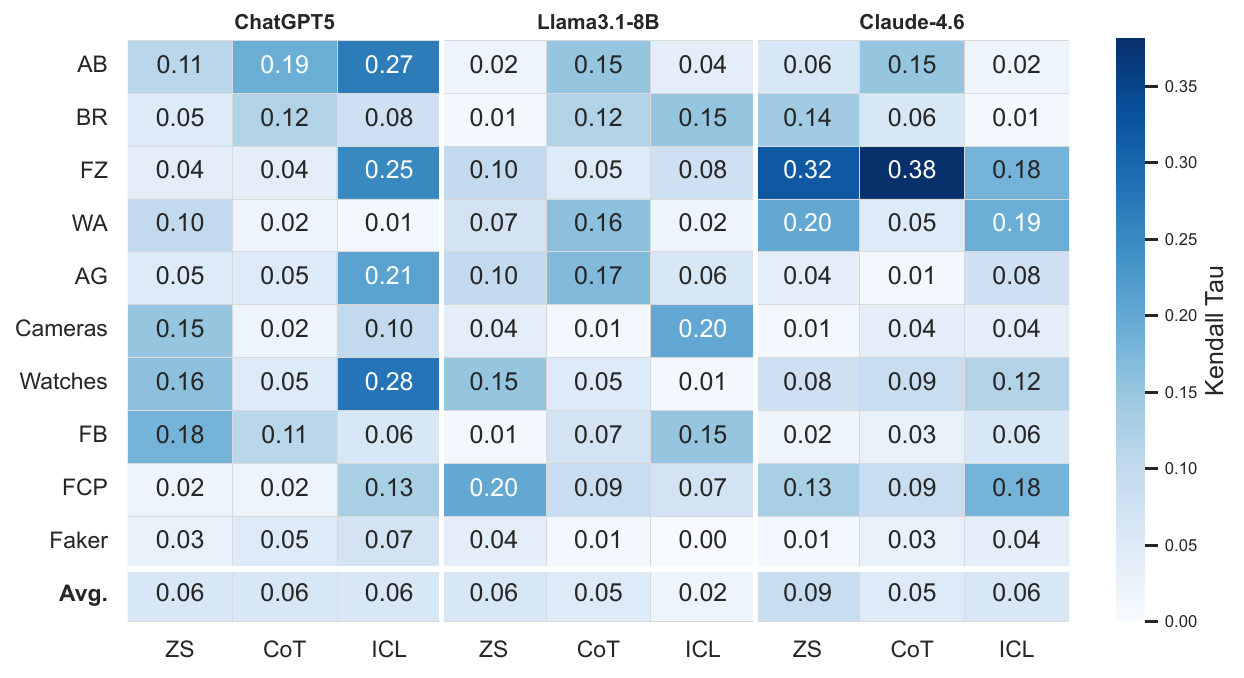}
}
\setlength{\tabcolsep}{2pt}
\subfigure[\normalsize{Cosine similarity between attribute and token-level counterfactual self-explanations ($\uparrow$).}]{
\includegraphics[width=\columnwidth]{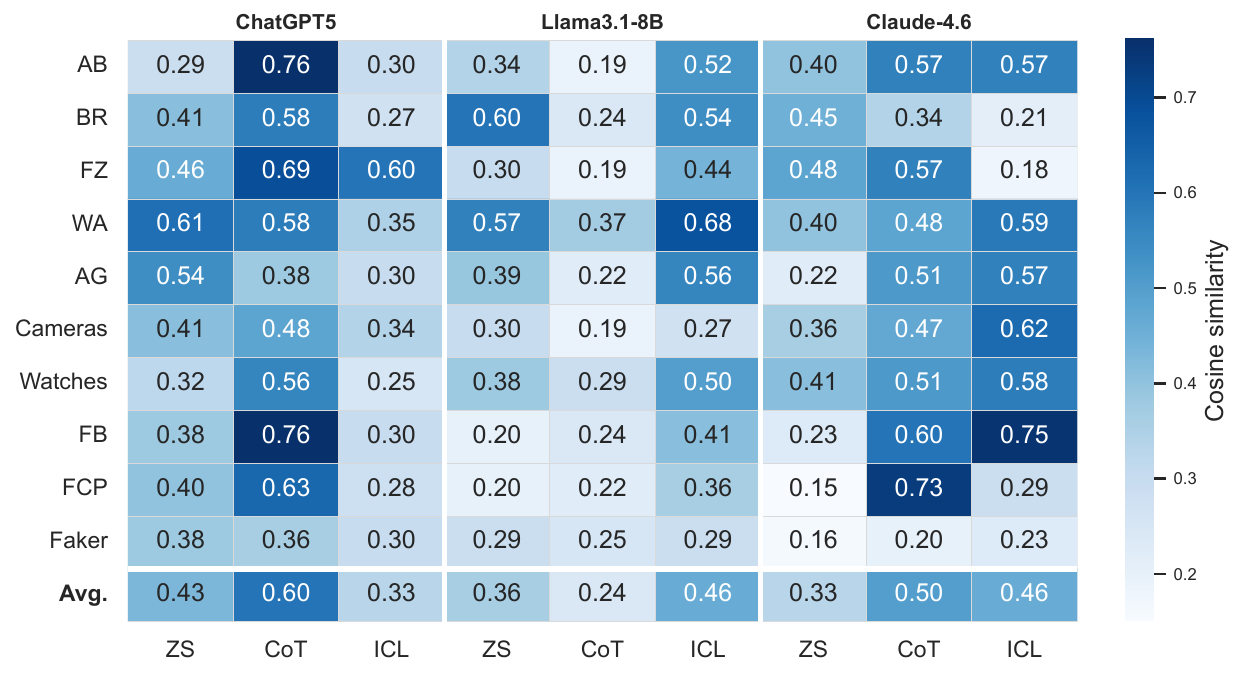}
}
\caption{Cross granularity consistency of self-explanations.}
\Description{Cross granularity consistency of self-explanations.}    
\label{fig:attribute-token-alignment}
\end{figure}

Figure~\ref{fig:attribute-token-alignment} reports the results of this analysis. Figure~\ref{fig:attribute-token-alignment}(a) shows that attribute-level and token-level feature attribution explanations are often misaligned: correlations are often close to zero, indicating negligible agreement. Across models, the highest, albeit still modest, correlation values are observed for ChatGPT-5 and Claude-4.6-Sonnet.
Figure~\ref{fig:attribute-token-alignment}(b) reports the cosine similarity scores for counterfactual explanations. In this case, a slightly stronger alignment between attribute-level and token-level explanations emerges. ChatGPT-5 achieves the largest alignment on average, while Claude-4.6-Sonnet shows alignment peaks for specific datasets (and prompts). 
However, across datasets, no consistent trend is observed.

\subsubsection{Alignment of feature attribution and Counterfactual explanations.}

To quantify the consistency between feature attribution and counterfactual explanations, we introduce a measure that captures the fraction of the total attribution score concentrated on the features modified by a counterfactual.

Let $x$ be an input record pair and let $\Delta(x)$ denote the set of features (attributes or tokens) modified by a counterfactual explanation for $x$. Let $s(f)$ be the attribution score assigned to feature $f$, and let $\mathcal{F}(x)$ denote the set of all features of $x$. We define the attribution score mass as
$
S(x) = \sum_{f \in \mathcal{F}(x)} s(f)$. We then define the \emph{Feature attribution-Counterfactual Alignment} (FCA) for input $x$ as:
$
\mathrm{FCA}(x) = \frac{\sum_{f \in \Delta(x)} s(f)}{S(x)}
$.
Intuitively, higher values of $\mathrm{FCA}(x)$ indicate stronger agreement between feature attribution and counterfactual explanations, as the features whose modification flips the prediction also account for a larger fraction of the model’s feature attribution.

As a complementary measure, we also evaluate the overlap between counterfactual modifications and the most important features identified by feature attribution explanations. Let $\mathrm{Top}\text{-}k(x) \subseteq \mathcal{F}(x)$ denote the set of the $k$ features with the highest attribution scores. We define the \emph{top-$k$ overlap} as
$\mathrm{TkO}(x) = 
\frac{|\Delta(x) \cap \mathrm{Top}\text{-}k(x)|}{|\Delta(x)|}$.
This metric measures the fraction of features modified by the counterfactual that are also among the top-$k$ most important features. Higher values indicate stronger alignment between feature attribution and counterfactual explanations, while lower values suggest that counterfactual changes affect features deemed less important by feature attribution explanations.

\begin{table}[t]
\centering
\caption{Evaluation metrics. $\Delta(x)$: features modified by the
  counterfactual; $s(f)$: feature attribution score;
  $S(x){=}\sum_{f \in \mathcal{F}(x)}\!s(f)$: total attribution mass.}
\label{tab:metrics}
\scriptsize
\setlength{\tabcolsep}{4pt}
\begin{tabular}{@{}lp{3.55cm}c@{}}
\toprule
\textbf{Metric} & \textbf{Definition} & \textbf{Dir.} \\
\midrule
Faithfulness
  & AUC of feature-occlusion curve over ranked attribution scores; measures
    whether attributed features causally drive predictions.
  & $\downarrow$ \\[3pt]
Validity
  & Fraction of counterfactuals that flip the prediction:
    $\frac{1}{|D|}\sum_x \mathbf{1}[\hat{y}(x') \neq \hat{y}(x)]$.
  & $\uparrow$ \\[3pt]
TkO
  & Top-$k$ overlap between counterfactual-modified and most-attributed
    features: $\frac{|\Delta(x)\cap\mathrm{Top\text{-}}k(x)|}{|\Delta(x)|}$.
  & $\uparrow$ \\[3pt]
FCA
  & Attribution mass on counterfactual-modified features:
    $\frac{\sum_{f\in\Delta(x)}s(f)}{S(x)}$.
  & $\uparrow$ \\
\bottomrule
\end{tabular}
\end{table}

The metrics used throughout the paper are summarize in Table~\ref{tab:metrics}.

\begin{figure}[t]
\subfigure[\normalsize{Attribute-level ($\uparrow$)}]{
\includegraphics[width=\columnwidth]{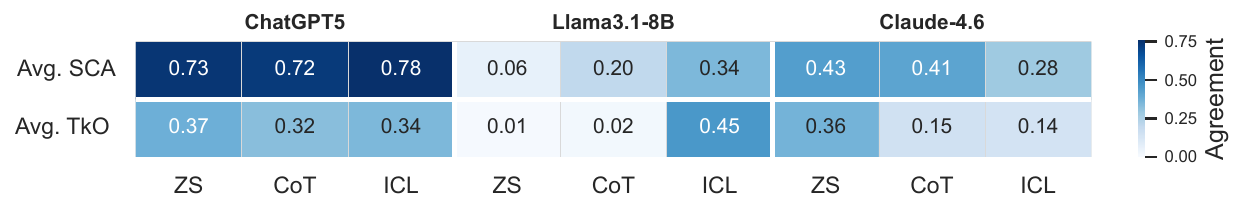}
}
\setlength{\tabcolsep}{3.7pt}
\subfigure[\normalsize{Token-level ($\uparrow$)}]{
\includegraphics[width=\columnwidth]{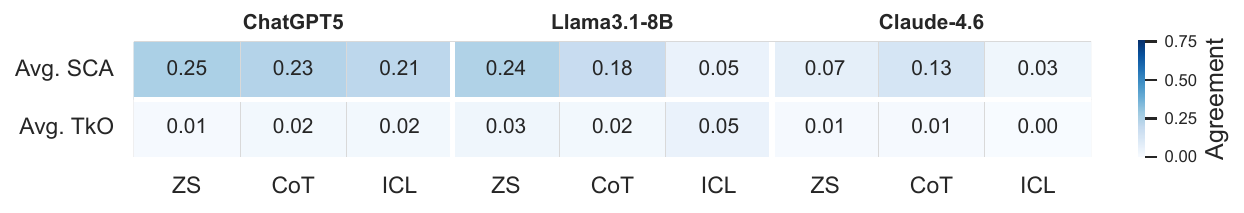}
}
\caption{Feature attribution and counterfactual agreement of self-explanations ($\uparrow$).}
\Description{Feature attribution and counterfactual agreement of self-explanations ($\uparrow$).}    
\label{fig:SCA}
\end{figure}

Figure~\ref{fig:SCA} reports the average FCA and Top-$k$ Overlap (TkO) aggregated across the considered datasets. For ChatGPT-5 and Llama-3.1 (especially), both metrics attain low values across prompting strategies, indicating limited agreement between feature attribution and counterfactual explanations. This suggests that, for those models, features receiving high attribution scores are not necessarily those whose modification is sufficient to alter the model’s prediction.

For Claude-4.6 we observe a higher degree of agreement between feature attribution and counterfactual explanations, across different prompts. 

At the attribute level, ICL improves alignment for Claude-4.6 and LLaMA~3.1–8B, leading to a higher concentration of attribution score on attributes modified by counterfactual explanations. In contrast, ChatGPT-5 achieves stronger alignment under ZS prompting, while CoT and ICL tend to degrade alignment, suggesting that explicit demonstrations or step-by-step reasoning may be less beneficial for stronger models.
At the token level, alignment further decreases in most configurations, likely due to the substantially larger feature space and more diffuse attribution score distributions. No specific prompt strategy seems to outperform the others, token-level explanations remain weakly aligned with counterfactual evidence across all models. These results suggest that prompting strategies can influence the stability of feature attribution explanations but do not consistently improve their alignment with counterfactual explanations, particularly at finer levels of granularity. On the other hand, different models might inherently produce  explanations with different degrees of alignment (but this won't necessarily translate into higher effectiveness, as reported in Figures~\ref{fig:self-explanation-Faithfulness} and \ref{fig:self-explanation-Counterfactual}).

\begin{figure}[t]
\subfigure[\normalsize{Attribute-level ($\downarrow$).}]{
\includegraphics[width=\columnwidth]{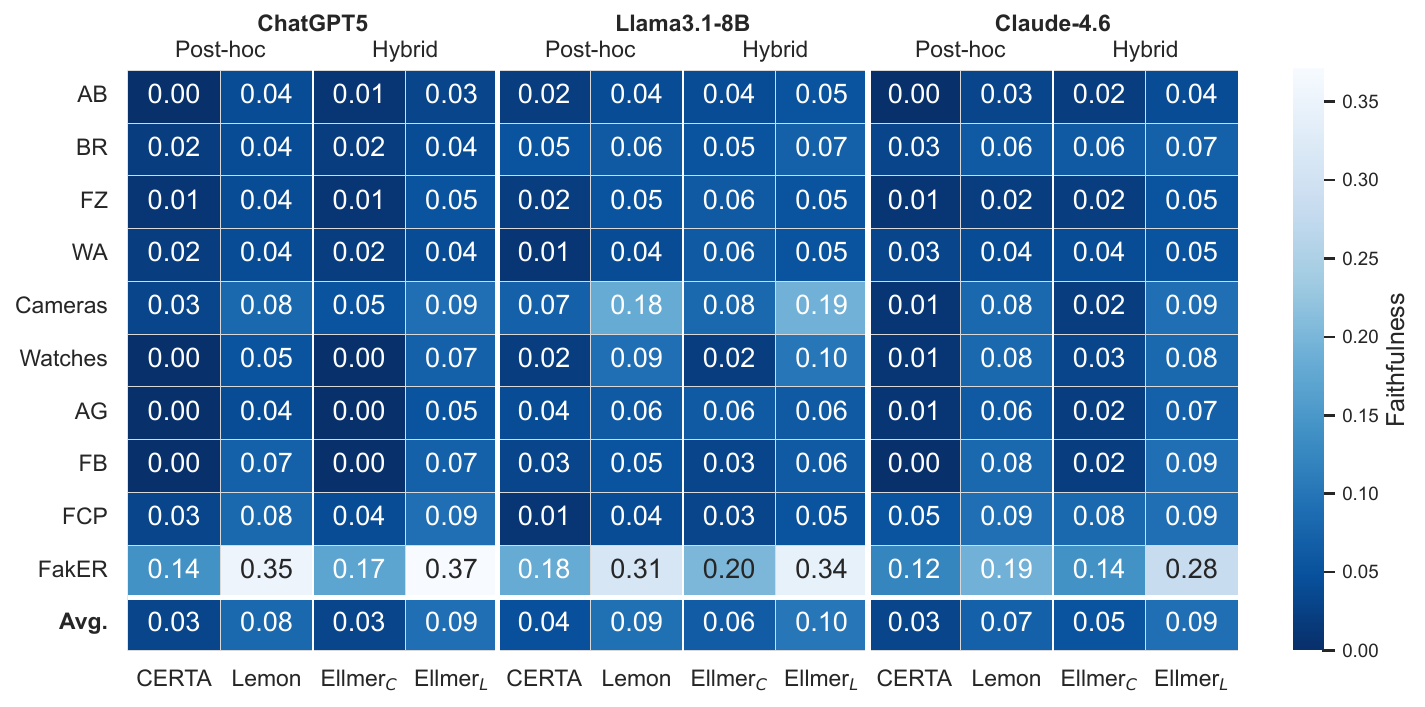}
}
\subfigure[\normalsize{Token-level ($\downarrow$).}]{
\includegraphics[width=\columnwidth]{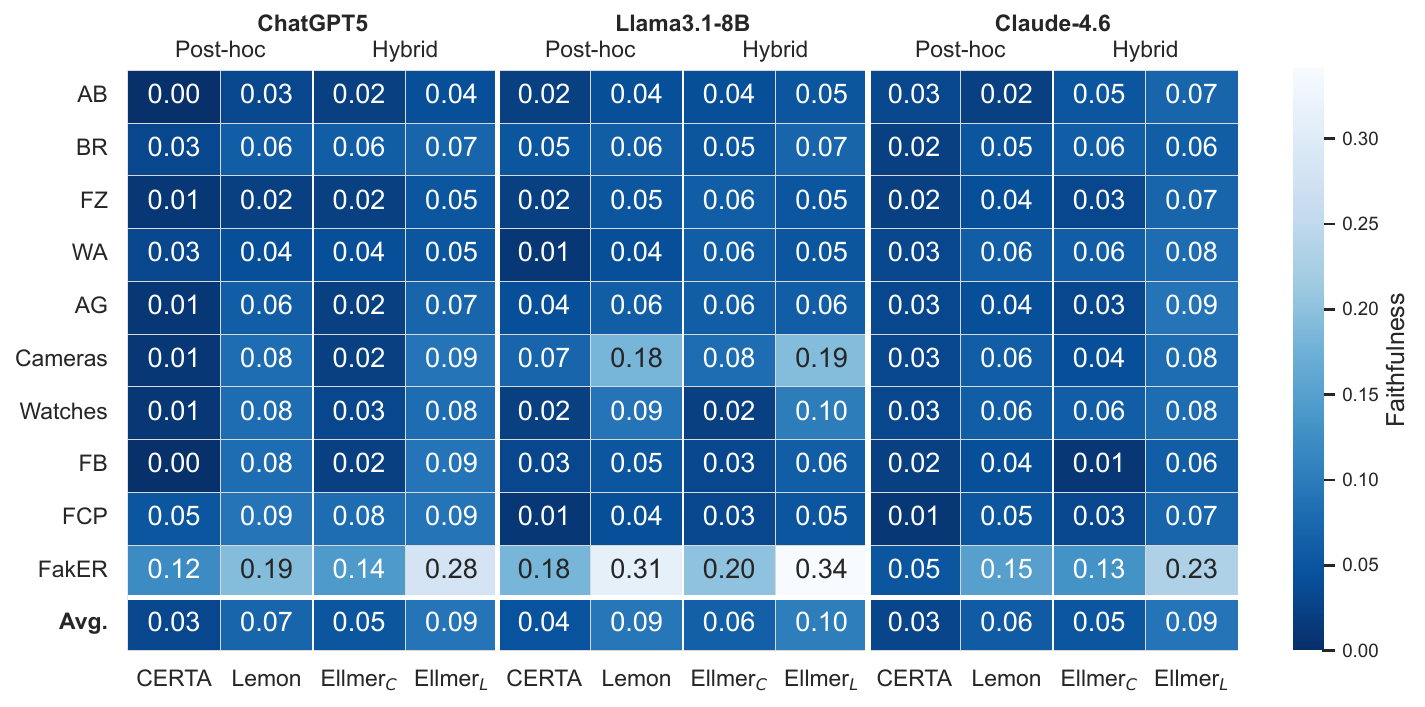}
}
\caption{Faithfulness of feature attribution explanations generated by post-hoc and hybrid methods. Values close to 0 (darker) are better ($\downarrow$). \uncerta{}$_{C}$ and  \uncerta{}$_{L}$ refer to instantiating {\uncerta} with the post-hoc method \certa{} and \lemon{}, respectively.}
\Description{Faithfulness of feature attribution post-hoc and hybrid explanations. Values close to 0 are better ($\downarrow$). \uncerta{}$_{C}$ and  \uncerta{}$_{L}$ refer to instantiating {\uncerta} with the post-hoc method \certa{} and \lemon{}, respectively.}
\label{tab:Faithfulness}
\end{figure}

\subsubsection{Probing the Underlying Explanation Mechanism}
\label{sec:introspection}
The faithfulness and validity scores reported above measure \emph{what} self-explanations produce, but they do not reveal \emph{how} an LLM arrives at its attribution scores or counterfactual proposals.
To investigate this, we designed an \emph{introspection} experiment: after we asked the LLM for the prediction and the feature attribution and counterfactual explanations, we extended the conversation with two targeted questions asking each model to describe in precise, formula-level terms (i) the computational procedure it used to assign attribution scores, and (ii) the strategy it followed to construct the counterfactual.
We collected 20 responses per model (10 from the books dataset and 10 from the carparts dataset) and analyzed them qualitatively.

\myparagraph{Claude-4.6-Sonnet.}
Claude consistently described a LIME-inspired \emph{heuristic}, but explicitly disclaimed that it had not executed the full LIME algorithm.
For feature attribution, it reported a pairwise attribute comparison rule: attributes on which the two records disagree receive a positive score, and attributes on which they agree receive a negative score, with magnitudes ranked according to domain-knowledge-based discriminative importance.
For counterfactuals, it described a greedy minimal-change procedure guided by its own attribution scores: target the highest attribution score conflicting attribute first, apply the minimal edit that would flip the disagreement, and stop as soon as the predicted label changes.
This behavior was stable across all 20 records. Claude's self-descriptions were the most \emph{transparent} of the three models: it neither claimed to run a formal XAI algorithm nor produced contradictory descriptions across instances.

\myparagraph{ChatGPT-5.}
ChatGPT-5 most often claimed to execute LIME formally, citing the proximity kernel $\pi_x(z) = \exp\!\bigl(-D(x,z)^2/\sigma^2\bigr)$, a weighted linear surrogate $g(z) = \beta_0 + \sum_i \beta_i z_i$, and identified its attribution scores with the learned coefficients $\beta_i$.
For counterfactuals, it described a feature attribution guided greedy single-attribute edit, analogous to Claude's strategy, preferring changes to title or plot fields for books and to name or part\_number fields for carparts.
However, several records contained explicit admissions that the values reported were ``illustrative placeholders'' or were derived from a ``rule-based heuristic'' rather than from an actual perturbation loop.
This inconsistency suggests that ChatGPT-5's formal LIME narrative functions as a post-hoc rationalization rather than a faithful account of the computation that actually took place.

\myparagraph{LLaMA-3.1-8B.}
LLaMA's responses were the least consistent.
The feature attribution mechanism cited varied substantially across records within the same dataset: word-embedding similarity, absolute attribute-value difference $|v_1 - v_2|$, token-frequency overlap, a LIME/SHAP variant, uniform-zero defaults, a dominance-based approach, and lexicographic distance all appeared in different responses.
The counterfactual descriptions were similarly erratic, with several records reporting bidirectional swaps (modifying both records simultaneously), erroneous field values, or simultaneous edits to all conflicting attributes rather than minimal single-attribute changes.

\myparagraph{Takeaway.}
None of the three models executed a formal XAI algorithm. The observed spectrum, from honest heuristic description (Claude) to claimed-but-unverified formal methods (ChatGPT-5) to incoherent behavior (LLaMA), mirrors and strenghtens the faithfulness gap reported above: the inability to reliably execute a principled computational procedure for feature attribution or counterfactual generation is a root cause of self-explanation unreliability for ER, independently of the prompting strategy used.
This finding reinforces the motivation for post-hoc methods, which, unlike self-explainers, execute deterministic algorithms whose attribution score and counterfactual constraints are computed outside the LLM.

\subsection{RQ-2: Are post-hoc explanations more trustworthy?}

To answer this research question, analogously to the analysis for RQ-1, we assess trustworthiness by measuring faithfulness for feature attribution, and validity for counterfactual explanations.

\subsubsection{Faithfulness and Validity of Self-Explanations}
Figure~\ref{tab:Faithfulness} reports faithfulness scores for post-hoc and hybrid explanation methods. Compared to self-explanations (previously reported in Figure~\ref{fig:self-explanation-Faithfulness}), post-hoc and hybrid approaches exhibit substantially higher trustworthiness, with faithfulness values consistently below $0.1$ across datasets and models, at both the attribute and token levels. In contrast, self-explanations yield higher (hence worst) faithfulness values, ranging from $0.15$ to $0.32$ at the attribute level, and from $0.13$ to $0.31$ at the token level.
Overall, the quality of the explanation is not influenced by the LLM, both at the attribute and token-level: for the same explanation method, the results provided by the three LLM are very close. 

\begin{figure}[t]
\subfigure[\normalsize{Attribute-level ($\uparrow$).}]{
\includegraphics[width=\columnwidth]{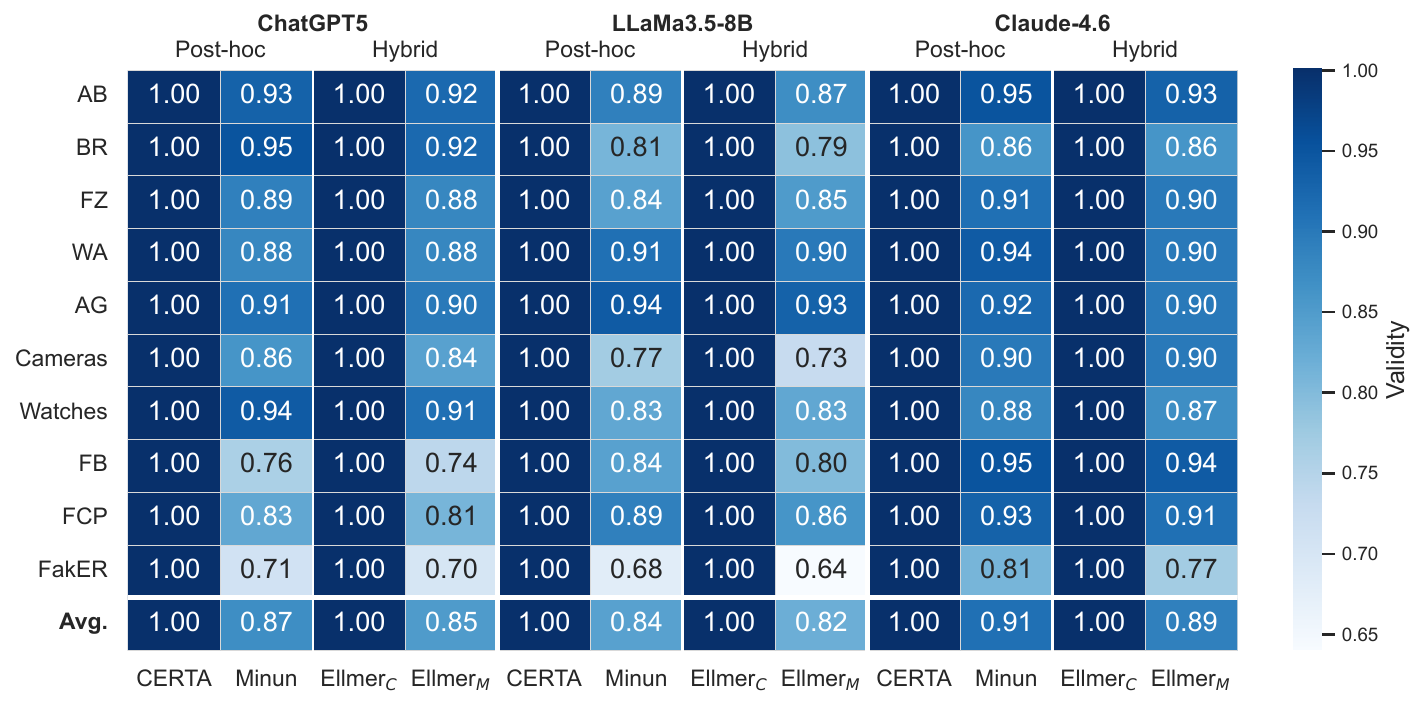}
}
\subfigure[\normalsize{Token-level ($\uparrow$).}]{
\includegraphics[width=\columnwidth]{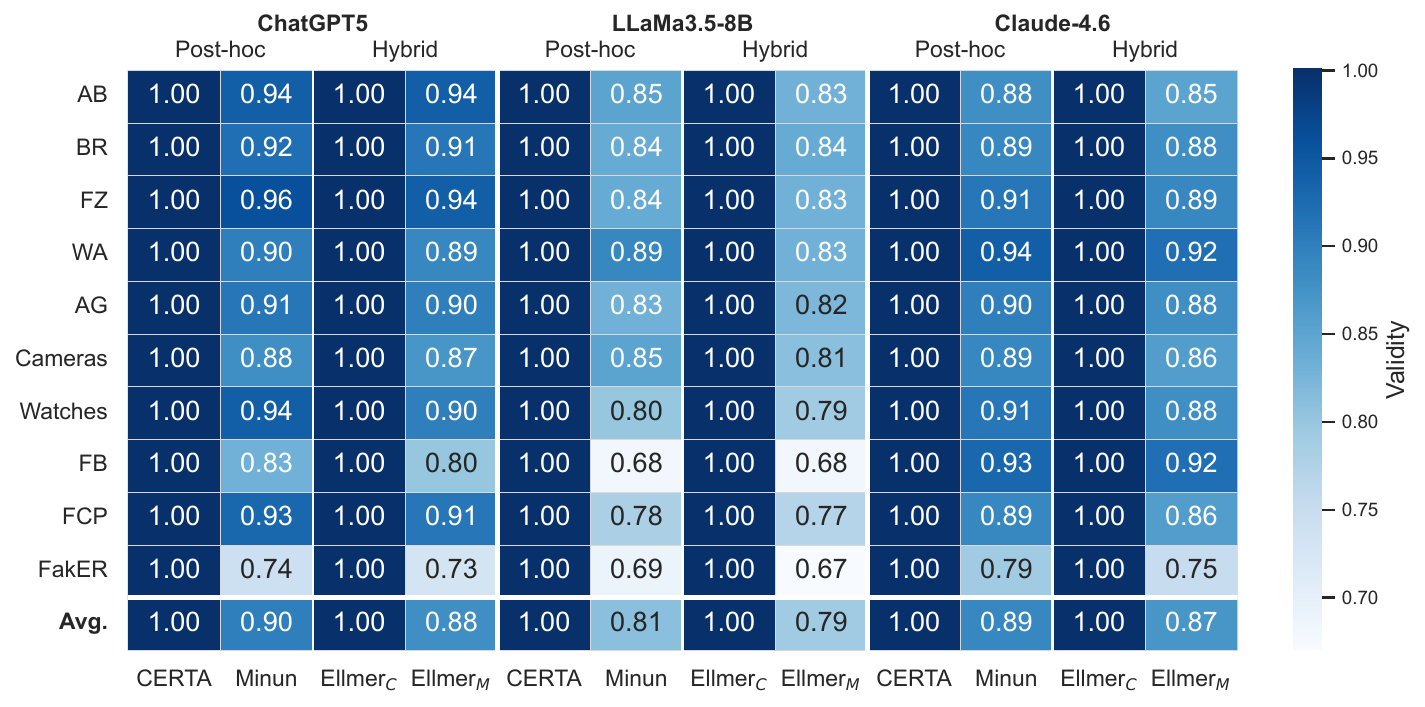}
}
\caption{Validity of counterfactual explanations. Values close to 1 (darker) are better ($\uparrow$). \uncerta{}$_{C}$ and  \uncerta{}$_{M}$ refer to instantiating {\uncerta} with the post-hoc method \certa{} and \minun{}, respectively.}
\Description{Validity of counterfactual explanations: (a) at attribute-level, (b) at token-level. Values close to 1 are better ($\uparrow$). \uncerta{}$_{C}$ and  \uncerta{}$_{M}$ refer to instantiating {\uncerta} with the post-hoc method \certa{} and \minun{}, respectively.}
    \label{tab:post-hoc-Validity}
\end{figure}

For counterfactual explanations, Figure~\ref{tab:post-hoc-Validity} reports the results of our assessment, showing that also in this case post-hoc and hybrid approaches consistently outperform self-explanations (in Figure~\ref{fig:self-explanation-Counterfactual}). At the attribute level, validity ranges from $0.82$ to $1$ for post-hoc and hybrid methods, compared to $0.63$–$0.81$ for self-explanations. At the token level, validity ranges from $0.67$ to $1$ for post-hoc and hybrid methods, while self-explanations exhibit comparable but generally lower performance. Notably, validity is consistently equal to $1$ across all datasets and models when using \certa{} and \uncerta{}$_C$.

In contrast to self-explanations, post-hoc and hybrid explanations exhibit stable performance across datasets, with no marked degradation on token-dense records. This behavior suggests that post-hoc and hybrid approaches effectively decouple explanation quality from dataset complexity. The only exception is FakER, a stress-test dataset with extreme feature cardinality, where performance slightly degrades for feature attribution explanations, while counterfactual explanations remain comparable to those on the other datasets.

It is important to observe that the hybrid approaches, represented by the \uncerta{} variants, achieve trustworthiness levels comparable to their corresponding post-hoc methods. For feature attribution explanations (Figure~\ref{tab:Faithfulness}), the performance of \uncerta{}$_C$ closely matches that of \certa{}, with an average difference of $0.01$ across models, while the results obtained by \lemon{} are comparable to those of the hybrid variant \uncerta{}$_L$. Similarly, for counterfactual explanations (Figure~\ref{tab:post-hoc-Validity}), \uncerta{}$_C$ and \certa{} achieve identical performance, and \minun{} and \uncerta{}$_M$ exhibit comparable validity scores.

For both faithfulness and validity, all the pairwise comparisons between self-explainers and post-hoc method are statistically significant ($p < 0.05$), with post-hoc methods consistently achieving lower faithfulness and higher validity. Mean differences in faithfulness span $0.048$ (ChatGPT-5 \textsc{CoT} vs.\ \lemon{}, token level) to $0.247$ (LLaMA \textsc{ZS} vs.\ \certa{}, token level). Mean validity gaps range from $-0.057$ (Claude \textsc{ICL} vs.\ \minun{}, attribute level) to $-0.379$ (LLaMA \textsc{ZS} vs.\ \certa{}, token level), with \certa{} consistently showing larger gaps relative to \minun{}/\lemon{}.

\begin{figure}[t]
\subfigure[\normalsize{Kendall-Tau correlation between attribute and token-level post-hoc and hybrid explanations ($\uparrow$).}]{
\includegraphics[width=\columnwidth]{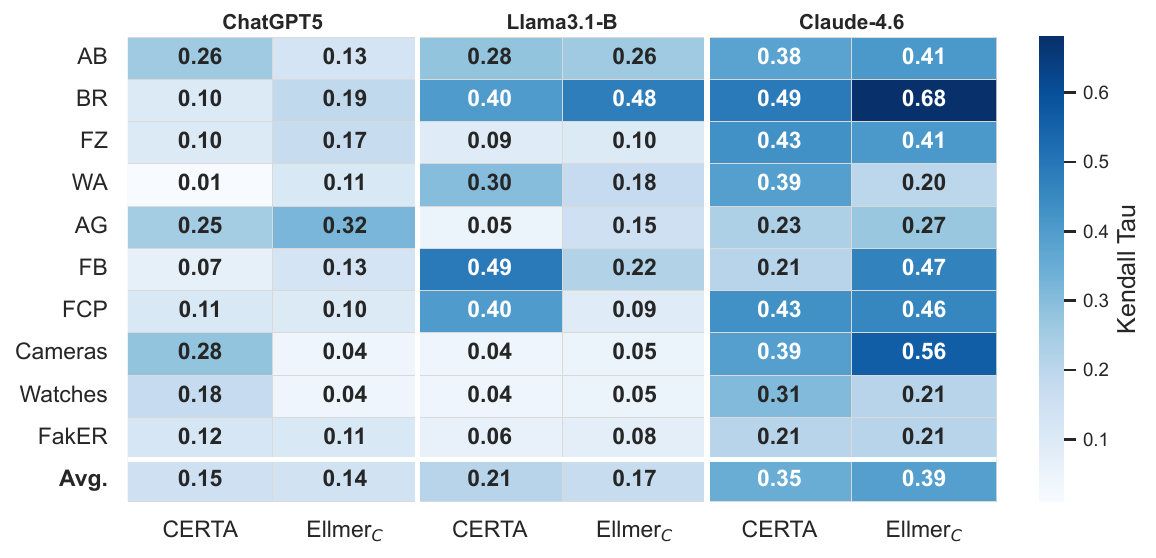}
}
\setlength{\tabcolsep}{3.7pt}
\subfigure[\normalsize{Cosine similarity between attribute and token-level counterfactual post-hoc and hybrid explanations ($\uparrow$).}]{
\includegraphics[width=\columnwidth]{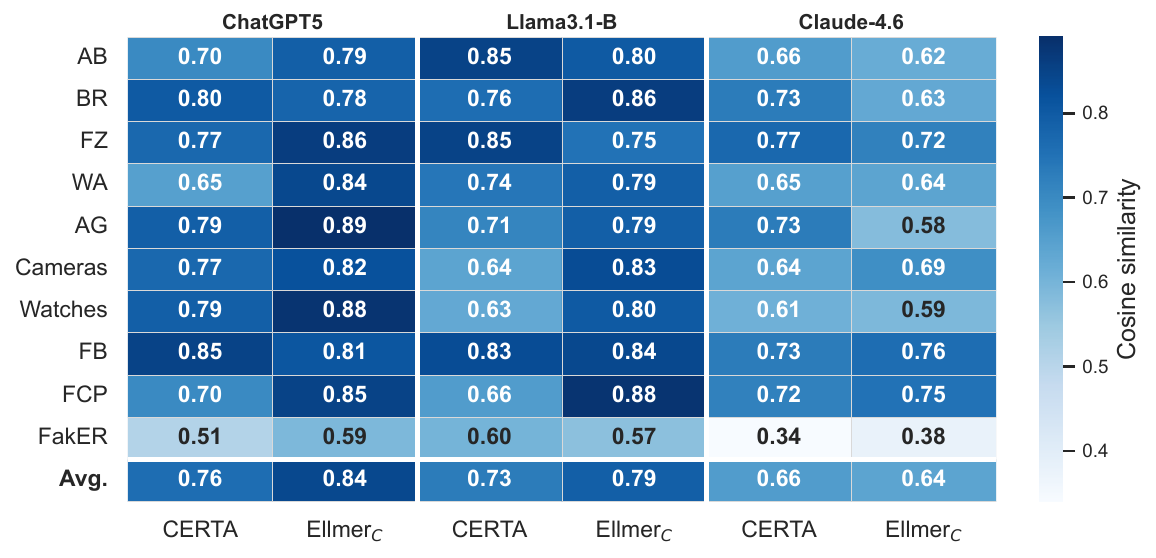}
}
\caption{Cross granularity consistency of post-hoc and hybrid explanations. Values close to 1 (darker) are better ($\uparrow$).}
\Description{Consistency of self explanations between attribute and token level ($\uparrow$).}    
\label{fig:attribute-token-alignment-post-hoc}
\end{figure}

\subsubsection{Cross-Granularity Consistency} 
Figure~\ref{fig:attribute-token-alignment-post-hoc} reports the cross-granularity consistency of post-hoc and hybrid explanations, measured by (a) the Kendall–Tau correlation between attribute-level and token-level feature attribution explanations, and (b) the cosine similarity between attribute-level and token-level counterfactual explanations.
Due to space constraints, we only report consistency results for \certa{} and {\uncerta{}$_C$}, which consistently achieved the strongest faithfulness and validity scores in preliminary experiments.
Across all models, datasets, and explanation methods, Kendall–Tau values are consistently positive but modest, indicating weak yet systematic agreement rather than random noise. On average, \certa{} achieves higher Kendall–Tau values than \uncerta{}$_C$ across all considered models; however, for ChatGPT-5 the difference is negligible, with the two methods exhibiting comparable average alignment.
The degree of feature attribution alignment varies notably across models and datasets. In particular, post-hoc explanations with LLaMA~3.1–8B show stronger alignment on smaller product datasets, whereas ChatGPT-5 achieves higher alignment on datasets with larger and richer records, such as WDC and FakER. Overall, both post-hoc and hybrid explanations consistently improve cross-granularity feature attribution alignment compared to self-explanations (Figure~\ref{fig:attribute-token-alignment}), especially for smaller models.
Turning to counterfactual explanations, cosine similarity values remain above $0.63$ across all models and datasets, with the exception of FakER, and average between $0.72$ and $0.84$. Hybrid explanations achieve slightly higher similarity than their post-hoc counterparts on average. The lower similarity observed for FakER is consistent with the substantially larger record size and token space of this dataset. With the exception of AG, counterfactual similarity values are markedly higher than those obtained for self-explanations, indicating a stronger and more stable cross-granularity alignment for post-hoc and hybrid methods.

Overall, these results show that post-hoc and hybrid explanations substantially improve cross-granularity consistency compared to self-explanations. While feature attribution alignment remains modest, it is systematic and non-random, and counterfactual explanations exhibit a much stronger and more stable agreement across levels of granularity. This further supports the effectiveness of hybrid approaches in combining reliability and efficiency without compromising explanatory coherence.

\begin{figure}[t]

\subfigure[\normalsize{Attribute-level ($\uparrow$)}]{
\includegraphics[width=\columnwidth]{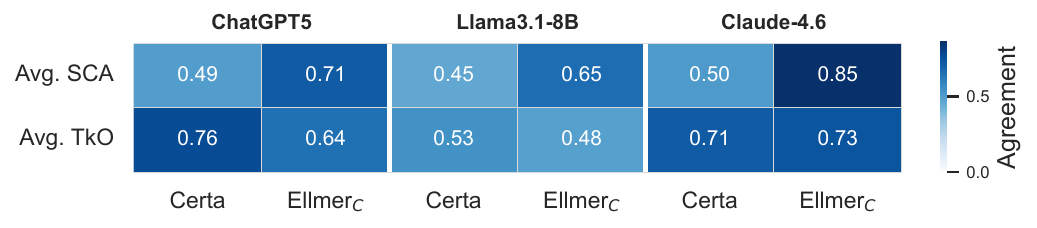}
}
\subfigure[\normalsize{Token-level ($\uparrow$)}]{
\includegraphics[width=.9\columnwidth]{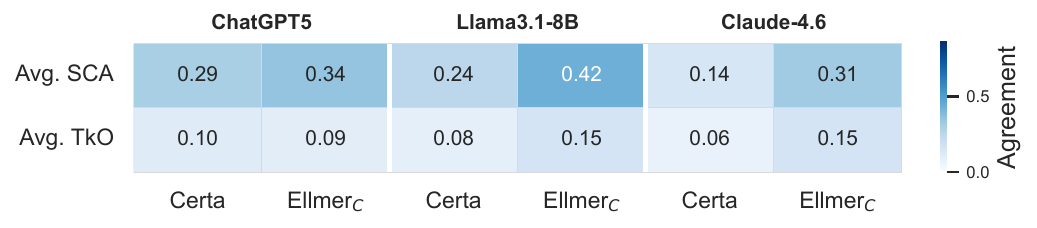}
}
\caption{Feature attribution and counterfactual agreement of post-hoc and hybrid explanations. Values close to 1 (darker) are better ($\uparrow$).}
\Description{Feature attribution and counterfactual agreement of of post-hoc and hybrid explanations. Values close to 1 (darker) are better ($\uparrow$).}    
\label{fig:SCA-post-hoc}
\end{figure}

\subsubsection{Feature attribution–Counterfactual Alignment}
Figure~\ref{fig:SCA-post-hoc} reports the agreement between feature attribution and counterfactual explanations for post-hoc and hybrid approaches, measured through FCA and Top-$k$ Overlap at both attribute and token levels. At the attribute level (Figure~\ref{fig:SCA-post-hoc}a), FCA values indicate that a substantial fraction of feature attribution scores is concentrated on attributes modified by counterfactual explanations. The hybrid approach \textsc{Ellmer}$_C$ consistently improves SCA over \certa{}, particularly for ChatGPT-5. Similarly, Top-$k$ Overlap reaches high values, showing that counterfactually modified attributes tend to appear among the most salient ones.
These results contrast those obtained for self-explanations (Figure~\ref{fig:SCA}), where feature attribution and counterfactual explanations exhibited weak or even negative alignment. This comparison highlights that post-hoc and hybrid methods induce a substantially more coherent relationship between attribution and counterfactual evidence.
At the token level (Figure~\ref{fig:SCA-post-hoc}b), both SCA and Top-$k$ Overlap are lower than at the attribute level, reflecting the larger feature space. Overall, these results indicate that post-hoc and hybrid approaches achieve a more reliable alignment between feature attribution and counterfactual explanations than self-explanations.

\subsection{RQ-3: Is it feasible to compute post-hoc explanations for
LLMs?} 
\label{sec:rq3}

Generating post-hoc explanations requires multiple predictions on samples created by perturbing the features of the original input. This is especially costly for token-level explanations, which demand significantly more predictions, increasing computational overhead.

\begin{figure}[t]
    \centering
    \includegraphics[width=0.95\columnwidth]{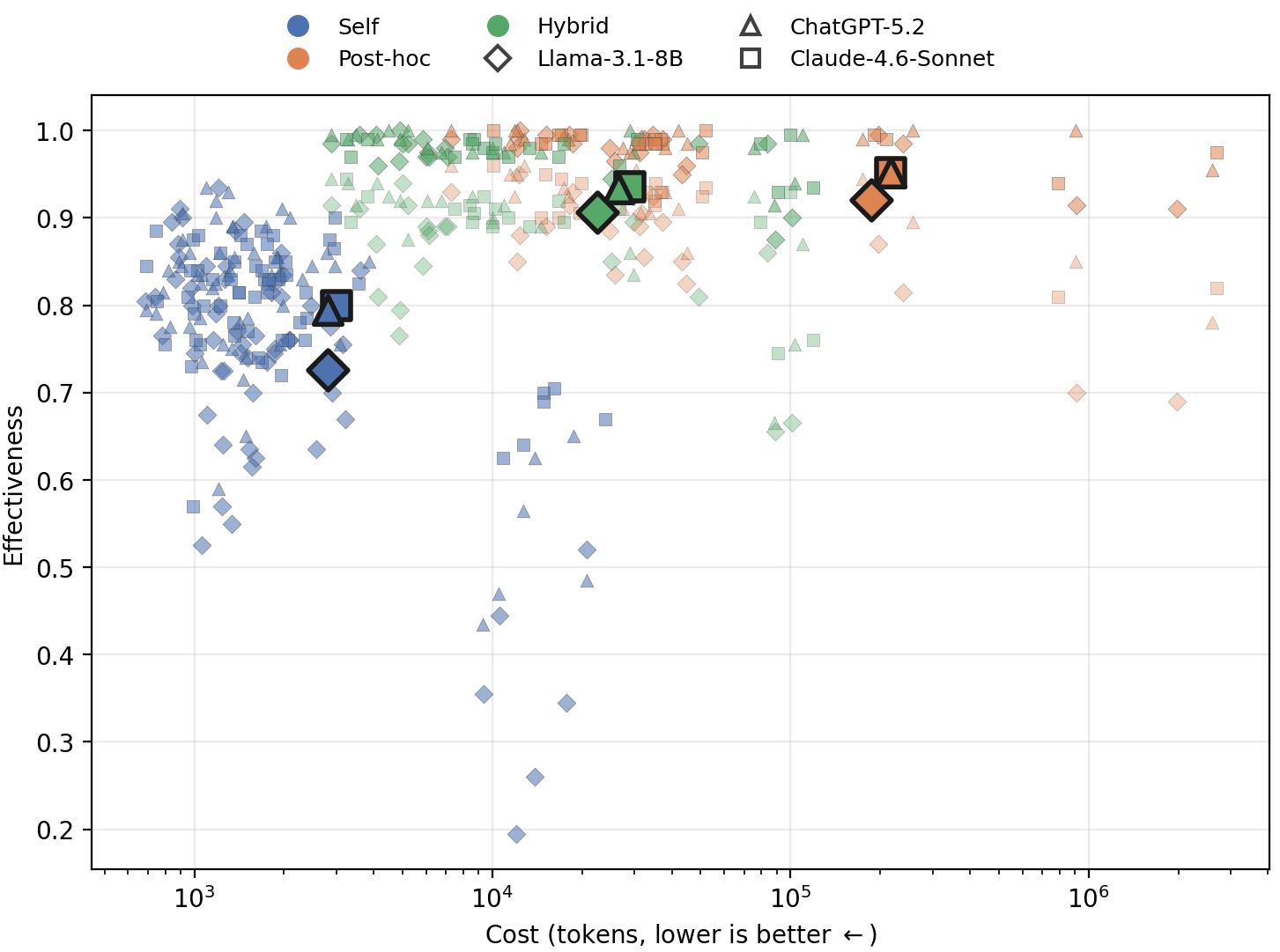}
    \caption{Effectiveness - Cost tradeoffs.}
    \Description{Effectiveness - Cost tradeoffs.}
    \label{fig:efficiency_effectiveness}
\end{figure}

Figure~\ref{fig:efficiency_effectiveness} summarizes the trade-off between explanation effectiveness and cost (expressed as number of tokens consumed by the LLM) across all explainers, models, and datasets. Three regions emerge clearly. Self-explainers occupy the low-cost regime but exhibit substantially lower and more variable effectiveness. Post-hoc methods achieve the highest effectiveness, but at significantly higher token cost. Hybrid \uncerta{} variants lie between these extremes, matching the effectiveness of post-hoc methods while requiring one to two orders of magnitude fewer tokens.
To validate these observations, we perform paired $t$-tests across all 60 experimental conditions. Both post-hoc and hybrid methods significantly outperform self-explainers in effectiveness (all $p<10^{-20}$), while post-hoc methods retain a small but significant advantage over hybrid methods. The opposite ordering emerges for cost: self-explainers are the cheapest, post-hoc methods the most expensive, and hybrid methods occupy an intermediate position, with all pairwise differences being significant ($p<10^{-21}$).
Overall, the results identify a clear Pareto frontier. Self-explainers minimize cost at the expense of explanation quality, whereas post-hoc methods maximize effectiveness but incur substantial computational overhead. Hybrid \uncerta{} variants provide the most favourable trade-off, recovering near post-hoc trustworthiness at a fraction of the cost and consistently bridging the gap between reliability and scalability across models and granularities.

\subsection{Correctness stratification}
\label{sec:correctness}

\begin{figure}[t]
    \centering
    \includegraphics[width=\columnwidth]{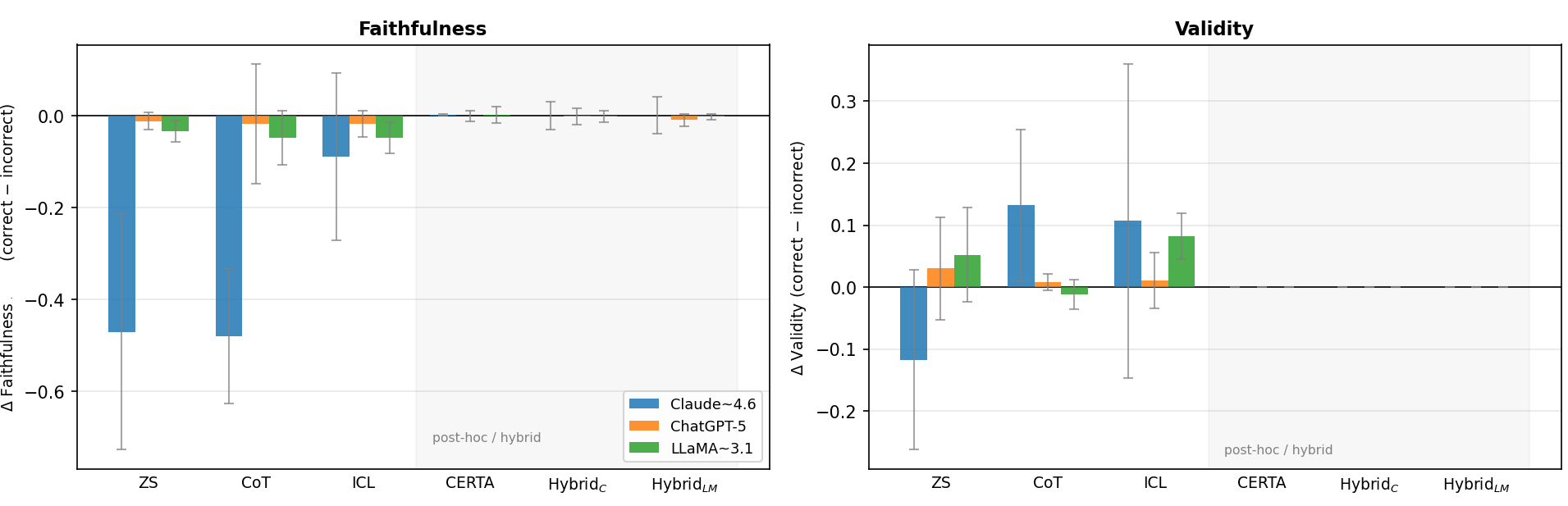}
    \caption{Mean difference in faithfulness  and validity between correct and incorrect predictions ($\Delta = \text{correct} - \text{incorrect}$) per explainer and model.}
    \Description{Mean difference in faithfulness  and validity between correct and incorrect predictions ($\Delta = \text{correct} - \text{incorrect}$) per explainer and model.}
    \label{fig:correctness_split_diff}
\end{figure}

Figure~\ref{fig:correctness_split_diff} reports the mean difference in faithfulness and validity between correct and incorrect predictions ($\Delta = \text{correct} - \text{incorrect}$) for each explainer and model. A negative $\Delta$ faithfulness indicates that explanations are more faithful when the underlying prediction is correct.
A pronounced effect emerges for Claude-4.6 self-explanations. Under \textsc{zs} and \textsc{cot}, faithfulness is substantially higher for correct than incorrect predictions ($\Delta=-0.471$ and $-0.480$, respectively), whereas ChatGPT-5 and LLaMA-3.1 remain close to zero (all within $[-0.05,0]$). \textsc{icl} largely mitigates the effect for Claude, reducing the gap to $-0.090$, suggesting that few-shot exemplars help anchor explanations to the input regardless of prediction outcome.
Validity shows a weaker pattern. Claude-4.6 exhibits moderate differences between correct and incorrect predictions (\textsc{zs}: $-0.117$, \textsc{cot}: $+0.132$, \textsc{icl}: $+0.107$), while ChatGPT-5 and LLaMA-3.1 again remain close to zero across prompting strategies.
In contrast, post-hoc and hybrid methods (\certa{}, \uncerta{}$C$, and \uncerta{}${LM}$) consistently exhibit $\Delta \approx 0$ for both metrics across all models, indicating that explanation quality is largely independent of prediction correctness. This invariance represents a qualitative advantage over self-explanation approaches and is only partially mitigated through improved prompting.

\subsection{Ellmer Sensitivity Analysis}
\label{sec:ellmer_sensitivity}
Both variants of \uncerta{} identify the same operating point, with $\phi=0.5$ providing the best quality--efficiency trade-off. \uncerta{}$_C$ is largely insensitive to both $\phi$ and model choice: validity remains close to $1.0$, faithfulness stays within a narrow range, and token consumption increases only modestly as $\phi$ grows. In contrast, \uncerta{}$_{LM}$ is more sensitive to both factors. As $\phi$ increases from $0.25$ to $0.5$, faithfulness decreases while validity improves; beyond $\phi=0.5$, both metrics largely plateau, providing little justification for better values despite the additional token cost. Overall, the sensitivity analysis indicates that $\phi=0.5$ offers the most favorable balance between explanation quality and efficiency across both variants. Finally, since the parameter $k$ (the initial number of top-salient features) is directly determined by $\phi$, both parameters exhibit consistent behavior.

\section{Related Work}
\label{sec:related}

LLMs are currently employed across a range of data management tasks~\cite{li2024llm,freire2025large}. Given their ability to contextualize the meaning of data, they can be considered an enabling technology especially in those tasks where the semantics of data plays a prominent role. 
\cite{DBLP:conf/adbis/PeetersB23} have shown the  potential of LLMs for the ER task.
On the other hand, despite their accuracy even in zero-shot and few-shot scenarios, they also exhibit a number of problematic behaviors, like not being able to respect the symmetric equality property (iff $A=B$ then $B=A$) as observed by \cite{DBLP:journals/corr/abs-2309-12288}.
The vulnerabilty of LLMs to adversarial prompts has been studied by~\cite{DBLP:journals/corr/abs-2306-04528}.

Our work falls in the broad area of interpretable AI~\cite{guidotti2018survey,lipton2018mythos} and causal inference~\cite{moraffah2020causal}.
Since understanding how LLMs inherently work is very hard~\cite{10.1145/3546577,zhao2023explainability,llm_selfexplain,srivastava2023beyond}, research is focusing  on the development of methods for interpreting and explaining LLMs, in different contexts and from different perspectives~\cite{zhao2023explainability,krishna2024post,10.1145/3546577}. 
\cite{llm_selfexplain} reported instability in LLM self-explanations but did not address how to reconcile efficiency with explanation reliability, which is the goal of our hybrid \uncerta{} approach.
\cite{turpin2023language} and \cite{wei2022chain} concentrated on 
CoT prompting, which provides improved performance along with implicit explanations by means of the illustrated step by step reasoning conducted by the LLM. However, \cite{turpin2023language} observed that such explanations are not always faithful. 
\cite{DBLP:journals/corr/abs-2307-13339} studied gradient-based feature attributions and showed that CoT can improve the robustness of token-level attribution scores to input perturbations.
\cite{DBLP:conf/acl/CifkaL23} studied how to quantify the importance of portions of the prompt with respect to the output, as the length of the prompt varies.

Post-hoc explanations have been proven effective in a variety of tasks~\cite{jesus2021can,madsen2022post}. 
As discussed in Section \ref{sec:posthoc}, several explanation systems have been proposed for the ER task~\cite{mojito,baraldi2021landmark,lemon,teofili2022effective,ebaid2019explainer,minun}. 

As we discussed along the paper, {\uncerta} is built on top of any post-hoc system based on perturbations and leverages the self-expanations of a LLM to overcome scalability issues.

\section{Conclusions}
\label{sec:conclusions}

This work empirically shows that LLM self-explanations for entity resolution are fluent and low-cost, yet exhibit notable trustworthiness limitations: relatively weak faithfulness, limited agreement between feature attribution and counterfactual explanations, and inconsistent cross-granularity behavior that prompting strategies alone do not fully address. Post-hoc methods offer substantially more reliable explanations, though at a considerable computational cost. \uncerta{} bridges this gap by using self-explanations to guide post-hoc exploration, preserving explanation quality while achieving meaningful reductions in runtime and token consumption.

An important open question left by this study concerns the \emph{root causes} of self-explanation unreliability.
Our introspection experiment (Section~\ref{sec:introspection}) offers a first qualitative window: none of the three models executed a principled attribution algorithm. 
The correctness stratification effect, where self-explanation faithfulness sometimes degrades for incorrect predictions (reported in Section~\ref{sec:correctness}) suggests models might be doing post-hoc rationalization. That may indicate that harder instances induce both prediction errors and less coherent feature attribution simultaneously.
Whether such behaviors reflect systematic artifacts in pretraining and instruction-tuning data~\cite{turpin2023language} or a more fundamental property of autoregressive self-explainers remains an open empirical question.
Investigating this through probing classifiers, causal  tracing~\cite{meng2022locating}, or controlled experiments that vary training data composition would likely yield actionable insights for the broader community working on LLM interpretability in structured prediction tasks.

\balance
\bibliographystyle{abbrv}
\bibliography{main}

\end{document}